\documentstyle[12pt,psfig]{ioplppt}
\def\n{{$n$}}

\def\etal{{\it et\thinspace al.}\ }

\def\eFe17{{\rm (e~+~Fe~\small XVII)}\ }
\def\fe16{{Fe$^{16+}$}}
\def\Fe17{{\rm Fe~{\small XVII}}\ }
\def\eie{{electron impact excitation}\ }
\def\cc{{close coupling}\ }
\def\bprm{{Breit-Pauli R-matrix}\ }
\def\apjl{{Astrophys. J. (Lett.)}}
\def\lr{{$\longrightarrow$}\ }

\def\sss{{\small SUPERSTRUCTURE}\ }
\newcommand{\abi}{{\it ab~initio}\ }

\newcommand{\ro}{{\rm o}}         

\begin{document}
\jl{2}
\title{Breit-Pauli R-matrix calculations for
electron impact excitation of Fe~XVII: a benchmark study}
[Electron impact excitation of Fe$^{16+}$]

\author{Guo-Xin Chen and Anil K. Pradhan}

\address{Department of Astronomy, 
Ohio State University \\ Columbus, Ohio,
USA, 43210}

\author{W. Eissner}
\address{Institut f\"ur Theoretische Physik, Teilinstitut 1, 70550 Stuttgart, Germany}

\begin{abstract}

A comprehensive study of relativistic and resonance effects in \eie of 
(e+Fe~XVII) is carried out using the \bprm (BPRM) method in the relativistic
\cc (CC) approximation. Two sets of eigenfunction expansions are employed; 
first, up to the $n$ = 3 complex corresponding 37 fine-structure levels (37CC) 
from 21 LS terms; second, up to the \n = 4 corresponding to 89 fine-structure 
levels (89CC) from 49 LS terms. In contrast to previous works,
the 37CC and the 89CC collision strengths exhibit considerable
differences. Denser and broader resonances due to \n = 4 are present in the 
89CC results both above and {\it below} the 37 thresholds, 
thus significantly affecting the collision strengths for the primary X-ray 
and EUV transitions within the first 37 \n = 3 levels. Extensive study of 
other effects on the collision strengths is also reported: (i) electric and 
magnetic multipole transitions E1,E2,E3 
and M1,M2, (ii) J-partial wave convergence of dipole and non-dipole
transitions, (iii) high energy behaviour compared to other
approximations. Theortical results are benchmarked against experiments to 
resolve longstanding
discrepancies --- collision strengths for the three prominent {\small X}-ray 
lines 3C, 3D and 3E at 15.014, 15.265, and 15.456~\AA\ 
are in good agreement with two independent 
measurements on Electron-Beam-Ion-Traps (EBIT). Finally, line ratios from a
collisional-radiative model using the new collisional rates
are compared with observations from stellar coronae and EBITs to illustrate 
potential applications in laboratory and astrophysical plasmas.
 
\end{abstract}

\pacno{34.80.Kw,32.30.Rj,95.30.Ky}
\maketitle

\submitted

\section{Introduction}

The ground configuration of \Fe17 has a stable
closed L-shell structure (neon core) rendering \Fe17 the dominant Fe ion 
species in many laboratory and cosmic plasmas. 
Owing to its importance \Fe17 has also become a benchmark ion, and one of 
the most extensively studied for its spectral diagnostic potential in 
plasma- and astro-physics.
Both the atomic structure of \Fe17 and \eie (EIE) of \eFe17 are extremely complex.
Although EIE of \Fe17
has been investigated experimentally and theoretically for a long time
\cite{bh92},
%since the Coulomb-Born calculations of Bely and Bely more than three decades ago \cite{bel}, 
there are considerable uncertainties in line intensity
comparisons between measurements and theories, even with the most
elaborate methods.
 
Unlike lower ionization stages of Fe, Fe~XVII is a medium-to-highly 
charged many-electron
ion with strong correlation and relativistic effects in both the 
target dynamics and
the electron collision processes \cite{ch99}, particularly for 
non-dipole forbidden and intercombination transitions and near-threshold
cross sections.
Both intermdediate coupling LSJ, and jj-coupling, are needed in atomic 
structure calculations 
for proper target level designations.
Precise atomic structure calculation is a crucial part of
EIE calculations. However, despite a number of experimental and theoretical
works, accurate atomic structure information on \Fe17 is not available in 
literature. The first 27 levels of \Fe17
in the configurations 2p$^5$3l (l=s,p,d) have been determined experimentally.
But consistent energies for inner-shell excitation
configurations 2s2p$^6$3l (l=s,p,d) (from level 28 to 37) are yet to be 
obtained. The same is true 
for the \n = 4 configurations (2p$^5$4l (l=s,p,d,f); 2s2p$^6$4l (l=s,p,d,f)).
Systematic theoretical structure calculations have been used to determine 
energies not available from experimental measurements, with an estimated
accuracy of less than a few to ten per cent, and an uncertainty of
$\sim$0.1$\AA$ in wavelengths to the ground state. 

Transition probabilities and \eie collision strengths
of \Fe17 are more inaccurate. We calculate transition probabilities
using the code \sss, which provides the structure input for
the EIE calculations of \Fe17 \cite{ei74}, and 
the  multi-configuration Dirac-Fock (MCDF) method using
the {\small GRASP} code with extensive configuration-expansion \cite{dy89}.
All E1, M1, E2 transitions among the 89 levels are calculated using
\sss and {\small GRASP} codes. We find some large M2 and E3 A-values from the 
comprehensive {\small GRASP} data, and for the first time, point out their importance 
in spectral formation of \Fe17.

Similar to the study of the \Fe17 structure, there is a long history of 
EIE calculations for \Fe17. Basically, there are three types of EIE 
calculations in literature.
(i) Distorted-wave (DW) and relativistic distorted-wave
(RDW), which couple initial and final channels
for a given transition and thus only background cross sections are calculated.
(ii) Isolated resonance approximation plus (R)DW: including a limited
number of resonances in the isolated resonance approximation (IRA).
This approach improves (i) somewhat;  however we show that the final
results are not very accurate due to the inevitably
limited number of channels in the IRA, and breakdown of IRA due to 
overlapping of dense and broad resonances in \Fe17 \cite{ch02}.
In some cases, the results from
this approach may be misleading or confusing owing to unknown missing 
channels.  
(iii) Close coupling (CC) approach: non-relativistic R-matrix calculations 
in LS coupling, and relativistic BPRM calculations in intermediate coupling. 
 While this is the most advanced method, and in principle capable of
including all important atomic effects, several issues require extreme
care, as demonstrated in this work.
Interestingly, there is
no complete R-matrix calculations in literature
for \Fe17 either. One previous coupled-channel calculation
by Mohan \etal \cite{mo97} employed
15 LS terms, followed by a pair coupling 
transformation to obtain fine-structure collision strengths between
levels in the first 27 levels. However, Mohan \etal obtained only
the background 
collision strengths, similar to the various DW calculations, because their
calculations were above the highest threshold and therefore no resonances
were included. Nontheless their data have been used to compare with
experiments and observations, and the conclusion that resonance enhancement 
is not important has been drawn.
Recently, limited BPRM calculations for \Fe17 were reported \cite{gu00}. Similar to
the work in \cite{mo97}, only 27CC was employed in the calculations. Also, no
detailed resonance structures in collision strengths were presented in this work
either and thus it prevents the in-depth investigations. 
In order to address these and other issues
thoroughly, it is therefore necessary to carry out a full set of 
BPRM calculations for \eFe17. 
There are some R-matrix calculations for the neon isoelectronic sequence
for other ions. Ne-like selenium has been calculated by fully
relativistic Dirac R-matrix approach \cite{wi91} and BPRM approach \cite{gu89}
using a 27CC expansion in jj-coupling and in LSJ coupling, respectively.

In the present work, relativistic and resonance effects are considered
with two sets of calculations: 
21 terms or 37 fine-structure levels (37CC) up to the \n = 3 complex, and
49 terms or 89 fine-structure levels (89CC) including up to the \n = 4
levels. The 37CC BPRM calculation includes
more than 7300 channels, with the largest Hamiltonian matrix of
dimension 3659; 
the 89CC BPRM calculation involves more than 20,000 channels overall,
with and the largest
the Hamiltonian matrix of dimension 10086 (this is possibly the largest
%the Hamiltonian matrix of dimension 10286 (this is possibly the largest
BPRM calculation yet carried out).
There are pronounced differences between the two sets of calculations.
The results show that the 89CC calculation with levels up to \n = 4 is 
necessary to obtain
accurate collision strengths even for the transitions up to the \n = 3,
since the \n = 4 resonances appear both above and, surprisingly,  {\it below}
the \n = 3 levels and affect the effective collision strengths
significantly.
The two sets of calculations also show that the backgrounds for some 
important transitions are also affected.
The implications of this study on \Fe17 for Ne-like ions and other
highly charged ions are pointed out.

 As mentioned, in addition to theoretical studies there have been several 
experimental measurements for \Fe17.
For example, experimental line intensity
ratios have recently been measured on Electron-Beam-Ion-Traps (EBIT) at Lawrence Livermore National
Laboratory (LLNL, Brown \etal 1998), and at the National Institute for
Standards and Technology (NIST, Laming \etal 2000), for
some prominent and strong lines of \Fe17 that have been observed 
in the {\small X}-ray spectra
of solar corona and other stellar coronae, active galactic nuclei,
{\small X}-ray binaries, supernovae, and recently in solar-type star
Capella from
{\small X}-ray satellites Chandra and XMM-Newton \cite{sa99}.
These 3 lines correspond to (see Fig.~1)
{\small X}-ray lines from M-shell 3d-2p with wavelength at 
$\sim$15$\AA$:
(1) 3C $\lambda$~15.013$\AA$:
1s$^2$2s$^2$2p$^5$[1/2]3d$_{3/2}$~$^1$P$^\ro_1$
(level 27)\lr1s$^2$2s$^2$2p$^6$~$^1$S$_0$ (ground state);
(2) 3D $\lambda$~15.265$\AA$:
1s$^2$2s$^2$2p$^5$[3/2]3d$_{5/2}$~$^3$D$^\ro_1$ (level
23)\lr1s$^2$2s$^2$2p$^6$~$^1$S$_0$;
(3) 3E $\lambda$~15.456$\AA$:
1s$^2$2s$^2$2p$^5$[3/2]3d$_{5/2}$~$^3$P$^\ro_1$ (level
17)\lr1s$^2$2s$^2$2p$^6$~$^1$S$_0$.
The line intensities display subtle effects since the 3C is a 
dipole-allowed transition,
but the 3D and 3E lines are intercombination transitions
that behave as
forbidden transitions (spin forbidden) at low-Z neon-like ions, 
and as allowed (in jj coupling) E1 transitions for high-Z high ions.
Comparison of experimental and theoretical data
shows disagreement of up to 50\% for 
 R1=3C/3D, and a factor of two for
R2=3E/3C \cite{br98,la00}. This means that some atomic mechanisms, and
their effect on line formation, have not been considered in previous 
theoretical works.

 The outline of the paper is as follows. In Secs.~2 and 3 the basic 
theoretical  and computational methods and techniques are briefly
described.
A schematic illustration of many important \Fe17 lines are shown
in Fig.~1, and discussed in Sec.~4.
In Secs.~5 and 6 we present the results for
cross sections and line intensities, and demonstrate
that: (i) dense resonance structure appeared in
all transitions over the entire energy range below the highest threshold 
in the 89CC BPRM collision strengths; in particular,
resonance enhancement generally dominates
forbidden and intercombination
transitions (but has not heretofore been studied), (ii)
the theoretical line intensity ratios for intercombination transitions, over
the strongest dipole-allowed transition,
agree with two sets of recent EBIT measurements \cite{br98,la00} 
to 10\% or within experimental uncertainties.
The conclusions are summarised in Sec.~7.

\section{Theory}

The \Fe 17 target wavefunctions are computed using \sss (Eissner \etal
1974 \cite{ei74}), which employs a scaled Thomas-Fermi-Dirac-Amaldi potential
(Eissner and Nussbaumer 1969 \cite{ei69}) to compute the set of one-electron orbitals.
The scaling parameters are optimised with a list of target
configurations and the Breit-Pauli Hamiltonian \cite{be95}. %(Berrington \etal 1995). 
In the Breit-Pauli R-matrix (BPRM) approximation \cite{sc82} the following Hamiltonian
terms are retained
\begin{eqnarray}
H_{N+1}^{\rm BP}=H_{N+1}+H_{N+1}^{\rm mass} + H_{N+1}^{\rm Dar}
+ H_{N+1}^{\rm so},
\end{eqnarray}
where $H_{N+1}$ is the non-relativistic Hamiltonian
together with the one-body mass correction term,
the Darwin term and the spin-orbit term
resulting from the reduction of the Dirac equation to the Pauli form.
The mass-correction
and Darwin terms do not break the LS symmetry, and they can therefore
be retained with a great effect in computationally cheaper LS
calculations. Spin-orbit interaction does, however, split the LS
{\em terms} into {\em fine-structure levels} labelled by $J{\mit\pi}$,
where J is the total angular momentum and $\pi$ the parity.                     

In the coupled channel or close coupling (CC) approximation
the wave function expansion,
$\Psi(E)$, for a total spin and angular symmetry  $SL\pi$ or $J\pi$,
of the (N+1)-electron system
is represented in terms of the target ion states as:
 
\begin{equation}
\Psi(E) = A \sum_{i} \chi_{i}\theta_{i} + \sum_{j} c_{j} \Phi_{j},
\end{equation}
 
\noindent
where $\chi_{i}$ is the target ion wave function in a specific state
$S_iL_i\pi_i$ or level $J_i\pi_i$, and $\theta_{i}$ is the wave function
for
the (N+1)$^{th}$ electron in a channel labeled as
$S_iL_i(J_i)\pi_i \ k_{i}^{2}\ell_i(SL\pi) \ [J\pi]$; $k_{i}^{2}$ is the
incident kinetic energy. In the second sum the $\Phi_j$'s are
correlation
wave functions of the (N+1)-electron system that (a) compensate for the
orthogonality conditions between the continuum and the bound orbitals,
and (b) represent additional short-range correlations that are often of
crucial importance in scattering and radiative CC calculations for each
$SL\pi$. The $\Phi_j$'s are also referred to as ``bound channels", as
opposed to the continuum or ``free" channels in the first sum over the
target states. In the relativistic BPRM calculations the set of ${SL\pi}$
are recoupled to obtain (e + ion) states  with total $J\pi$, followed by
diagonalization of the (N+1)-electron Hamiltonian. Details of the
diagonalization and the R-matrix method are given in many previous works
(e.g. Berrington \etal 1995).
 
\section{Computations}

\subsection{Target eigenfunctions}

The configuration-expansion consists of 49 LS terms
corresponding to 89 fine-structure levels with principal quantum number up to 
\n = 4. Target energies are given in Table 1 and compared with observed
ones. Based on both \sss and {\small GRASP} calculations we have assigned both
jj-coupling and intermediate coupling LSJ designations to the energy
levels. 
The MCDF {\small GRASP} calculations are with the
same configuration-expansion as used in the \sss calculations. The
MCDF method is a
complete self-consistent-field (SCF) procedure, which means both the orbitals
 and the expansion coefficients are variational. However, the Breit 
interaction is included
as a perturbation in the CI-type \sss calculations with fixed orbitals.
Selected oscillator strengths for the transitions among target
states are compared with other works in Table 2. We also report on some
higher multipole transition probabilities, compared to E1, in Table 3.
The extensive calculations on \Fe17 atomic structure will be reported
separately.

\subsection{Electron impact excitation of \ \Fe17}

 Target configurations
$$
2s^22p^6 ,\ 2s^22p^5 (3s,3p,3d),\ 2s^22p^5 (4s,4p,4d,4f),
\ 2s^12p^6 (3s,3p,3d),\ 2s^12p^6 (4s,4p,4d,4f)
$$
corresponding to 89 fine-structure target levels up to \n = 4 are included 
in the CC expansion. The largest symmetry 
%is total J = 7/2 (odd parity) with
%395 free channels and 486 bound channels ($\Phi_j$); the dimension of
%the Hamiltonian is 10286 with 25 continuum orbitals in the inner region. 
is total J = 9/2 (even parity) with
400 free channels and 86 bound channels ($\Phi_j$); the dimension of 
the Hamiltonian is 10086 with 25 continuum orbitals in the inner region.
The R-matrix boundary is at $R_0=3.969$ a.u.
The maximum electron impact energy is up to 400-500 Ry. BPRM collision strengths
are calculated for all J $\leq 51/2$, including all the partial waves in the
range $l=0-31$. This ensures convergence of the
collision strengths without partial wave top-up in the low-energy region. 
In the high-energy region, we use
relativistic distorted-wave (RDW), and/or
Coulomb-Born-Bethe (CBe), top-up procedures for the different types of 
transitions:
allowed, intercombination, forbidden, or other mixed or peculiar transitions.
 Special attention is paid to convergence with respect to resolution of
resonances. We use a constant-energy 
of up to 20,000 energies to compute
rate coefficients for practical applications.
 We also carry out a 37CC calculation with the \n = 3 levels
in order to check the accuracy of previous DW and other calculations,
and to demonstrate the necessity of performing 89CC calculations.
The other R-matrix parameters in the 37CC calculations were the same as the 89CC.

\section{Important \Fe17 lines}

To illustrate applications of the present \Fe17 results to astrophysical
objects and laser transtiions in plasmas, we show
selected {\small X}-ray transitions to the ground level, and
{\small X/UV}-ray transitions between excited levels,
in Figs.~1(a) and 1(b). 
We have computed 3653 transition probabilities and
collision strengths for 3916 transitions among the 89 levels
for line intensity modeling in a collisional-radiative model (CRM).
The 20 important lines are shown in Fig.~1(b).
Some of them,
dipole E1 transitions 3A (1-33) and 3B (1-31), E2S (1-37) transition from an upper
level with an inner-shell 2s hole, the strong E1 line 3C (1-27), and two 
intercombination transitions 3D (1-23) and 3E (1-17)
from 2p$^5$3d levels to the ground state, show prominent emission and/or absorption
features in collision ionized plasmas. These lines
have been extensively used in the 
diagnostics of temperature, density and ionization balance, 

Fig.~1(b) also shows E1 allowed and intercombination lines 3G (1-3) and 3F (1-5)
respectively, and a M2 (1-2) line,
from low-lying upper levels of the 2p$^5$3s configuration. 
The upper levels 3F and 3G have very fast radiative
decay rates. This point is very important in the creation
of population inversion of soft {\small X}-ray
laser lines in \Fe17. As opposed
to the high-lying lines from 2p$^5$3d levels and others, cascade effects
may dominate the line formation of 3F, 3G, and M2. These three low-lying
lines are powerful diagnostic tools used in photoionized plasmas.

An E2L (1-7) line from 2p$^5$3p has also been observed from EBIT experiments,
and possibly from some cosmic plasmas. The {\small X/UV}
line $\lambda$~1153~$\AA$ (3-4) between excited levels 3 and 4 is a potential
diagnostic of temperature, such as in solar flares,  by comparing its 
intensity to the $\lambda$~975~$\AA$ line in Fe~{\small XVIII}.

There are two important points related to the {\small X/UV}-ray lines 
in Fig.~1(b) among excited levels from 2p$^5$3p to 2p$^5$3s:
(1) fast radiative decay of levels 3 and 5, and
(2) strong {\it monopole} collisional excitation rate to level 15 
(2p$^5$3p $^1S_0$) responsible for
population inversion leading to the strongest soft {\small X}-ray laser lines.
In addition to the collisional excitation effects, cascade effects may
also contribute to the level population of level 15.

\section{Results and discussions}

\subsection{\Fe17 atomic structure}

{\tiny
\begin{table}
\caption{The 89 fine-structure $n=2$, 3 and 4 levels
included in the BPRM calculation and their calculated and
observed energies in Rydbergs for Fe~{\scriptsize XVII}. `obs' values are observed values from
NIST website ({\it http://www.nist.org}) (level denoted by `$\dag$' is from \cite{br98}.
%Brown \etal (1998)). 
`SS' and `MCDF' are from {\scriptsize SUPERSTRUCTURE} and {\scriptsize GRASP} calculations,
respectively. The index $i$ is used for transition keys. {NB:, both jj-coupling and LS-coupling
notations are needed for the level designations of  Fe~{\scriptsize XVII}.}
}
\begin{center}
\begin{tabular} {rlrrr|rlrrr}
\hline
\noalign{\smallskip}
 $i$&Level&obs&SS&MCDF&
 $i$&Level&obs&SS&MCDF\\
\noalign{\smallskip}
\hline
\noalign{\smallskip}
  1&2s$^2$2p$^6$\ $^1{\rm S}_0$\ (0,0)0&0.0&0.0&0.0&
 46&2s$^2$2p$^5$4p\ $^3{\rm P}_2$\ (3/2,3/2)2&&72.9701 &72.7671\\
  2&2s$^2$2p$^5$3s\ $^3{\rm P}_2^\ro$\ (3/2,1/2)$^\ro$2&53.2965&53.3542&53.1684&
 47&2s$^2$2p$^5$4p\ $^3{\rm P}_0$\ (3/2,3/2)0&&73.2625&73.0529\\
  3&2s$^2$2p$^5$3s\ $^1{\rm P}_1^\ro$\ (3/2,1/2)$^\ro$1&53.43&53.4982&53.3100&
 48&2s$^2$2p$^5$4p\ $^3{\rm D}_1$\ (1/2,1/2)1&&73.7627&73.5563\\
  4&2s$^2$2p$^5$3s\ $^3{\rm P}_0^\ro$\ (1/2,1/2)$^\ro$0&54.2268&54.2786&54.0957&
 49&2s$^2$2p$^5$4p\ $^3{\rm P}_1$\ (1/2,3/2)1&&73.8425&73.6430\\
  5&2s$^2$2p$^5$3s\ $^3{\rm P}_1^\ro$\ (1/2,1/2)$^\ro$1&54.3139&54.3726&54.1851&
 50&2s$^2$2p$^5$4p\ $^1{\rm D}_2$\ (1/2,3/2)2&&73.8597&73.6595\\
  6&2s$^2$2p$^5$3p\ $^3{\rm S}_1$\ (3/2,1/2)1&55.5217&55.5586&55.3963&
 51&2s$^2$2p$^5$4d\ $^3{\rm P}_0^\ro$\ (3/2,3/2)$^\ro$0&&74.0121&73.8044\\
  7&2s$^2$2p$^5$3p\ $^3{\rm D}_2$\ (3/2,1/2)2&55.7787&55.8321&55.6606&
 52&2s$^2$2p$^5$4d\ $^3{\rm P}_1^\ro$\ (3/2,3/2)$^\ro$1&73.95&74.0462&73.8369\\
  8&2s$^2$2p$^5$3p\ $^3{\rm D}_3$\ (3/2,3/2)3&55.8974&55.9429&55.7791&
 53&2s$^2$2p$^5$4d\ $^3{\rm F}_4^\ro$\ (3/2,5/2)$^\ro$4&&74.0921&73.8777\\
  9&2s$^2$2p$^5$3p\ $^1{\rm P}_1$\ (3/2,3/2)1&55.9804&56.0297&55.8654&
 54&2s$^2$2p$^5$4d\ $^3{\rm P}_2^\ro$\ (3/2,5/2)$^\ro$2&&74.1021&73.8904\\
 10&2s$^2$2p$^5$3p\ $^3{\rm P}_2$\ (3/2,3/2)2&56.1137&56.1586&56.9950&
 55&2s$^2$2p$^5$4p\ $^1{\rm S}_0$\ (1/2,1/2)$^\ro$0&&74.0943&73.9033\\
 11&2s$^2$2p$^5$3p\ $^3{\rm P}_0$\ (3/2,3/2)0&56.5155&56.5799&56.4050&
 56&2s$^2$2p$^5$4d\ $^3{\rm F}_3^\ro$\ (3/2,3/2)$^\ro$3&&74.1080&73.8994\\
 12&2s$^2$2p$^5$3p\ $^3{\rm D}_1$\ (1/2,1/2)1&56.6672&56.7209&56.5495&
 57&2s$^2$2p$^5$4d\ $^1{\rm D}_2^\ro$\ (3/2,3/2)$^\ro$2&&74.1534&73.9456\\
 13&2s$^2$2p$^5$3p\ $^3{\rm P}_1$\ (1/2,3/2)1&56.9060&56.9475&56.7855&
 58&2s$^2$2p$^5$4d\ $^3{\rm D}_3^\ro$\ (3/2,5/2)$^\ro$3&&74.1864&73.9736\\
 14&2s$^2$2p$^5$3p\ $^1{\rm D}_2$\ (1/2,3/2)2&56.9336&56.9798&56.8135&
 59&2s$^2$2p$^5$4d\ $^3{\rm D}_1^\ro$\ (3/2,5/2)$^\ro$1&74.30&74.3841&74.1666\\
 15&2s$^2$2p$^5$3p\ $^1{\rm S}_0$\ (1/2,1/2)0&57.8894&58.0832&57.9308&
 60&2s$^2$2p$^5$4f\ $^3{\rm D}_1$\ (3/2,5/2)1&&74.6730&74.4521\\
 16&2s$^2$2p$^5$3d\ $^3{\rm P}_0^\ro$\ (3/2,3/2)$^\ro$0&58.8982&58.9384&58.7738&
 61&2s$^2$2p$^5$4f\ $^1{\rm G}_4$\ (3/2,5/2)4&&74.6761&74.4522\\
 17&2s$^2$2p$^5$3d\ $^3{\rm P}_1^\ro$\ (3/2,3/2)$^\ro$1&58.981&59.0169&58.8454&
 62&2s$^2$2p$^5$4f\ $^3{\rm G}_5$\ (3/2,7/2)5&&74.6799&74.4535\\
 18&2s$^2$2p$^5$3d\ $^3{\rm P}_2^\ro$\ (3/2,5/2)$^\ro$2&59.0976&59.1644&58.9826&
 63&2s$^2$2p$^5$4f\ $^3{\rm D}_2$\ (3/2,7/2)2&&74.6855&74.4623\\
 19&2s$^2$2p$^5$3d\ $^3{\rm F}_4^\ro$\ (3/2,5/2)$^\ro$4&59.1041&59.1820&58.9901&
 64&2s$^2$2p$^5$4f\ $^3{\rm D}_3$\ (3/2,7/2)3&&74.7109&74.4871\\
 20&2s$^2$2p$^5$3d\ $^3{\rm F}_3^\ro$\ (3/2,3/2)$^\ro$3&59.1611&59.2240&59.0498&
 65&2s$^2$2p$^5$4f\ $^3{\rm F}_2$\ (3/2,5/2)2&&74.7141&74.4910\\
 21&2s$^2$2p$^5$3d\ $^1{\rm D}_2^\ro$\ (3/2,3/2)$^\ro$2&59.2875&59.3519&59.1797&
 66&2s$^2$2p$^5$4f\ $^1{\rm F}_3$\ (3/2,5/2)3&&74.7170&74.4935\\
 22&2s$^2$2p$^5$3d\ $^3{\rm D}_3^\ro$\ (3/2,5/2)$^\ro$3&59.3665&59.4492&59.2598&
 67&2s$^2$2p$^5$4f\ $^3{\rm F}_4$\ (3/2,7/2)4&&74.7246&74.4992\\
 23&2s$^2$2p$^5$3d\ $^3{\rm D}_1^\ro$\ (3/2,5/2)$^\ro$1&59.708&59.7904&59.6082&
 68&2s$^2$2p$^5$4d\ $^3{\rm F}_2^\ro$\ (1/2,3/2)$^\ro$2&&75.0232&74.8203\\
 24&2s$^2$2p$^5$3d\ $^3{\rm F}_2^\ro$\ (1/2,3/2)$^\ro$2&60.0876&60.1439&59.9749&
 69&2s$^2$2p$^5$4d\ $^3{\rm D}_2^\ro$\ (1/2,5/2)$^\ro$2&&75.0470&74.8391\\
 25&2s$^2$2p$^5$3d\ $^3{\rm D}_2^\ro$\ (1/2,5/2)$^\ro$2&60.1617&60.2190&60.0344&
 70&2s$^2$2p$^5$4d\ $^1{\rm F}_3^\ro$\ (1/2,5/2)$^\ro$3&&75.0704&74.8618\\
 26&2s$^2$2p$^5$3d\ $^1{\rm F}_3^\ro$\ (1/2,5/2)$^\ro$3&60.197&60.2643&60.0754&
 71&2s$^2$2p$^5$4d\ $^1{\rm P}_1^\ro$\ (1/2,3/2)$^\ro$1&75.17&75.2381&75.0263\\
 27&2s$^2$2p$^5$3d\ $^1{\rm P}_1^\ro$\ (1/2,3/2)$^\ro$1&60.6903&60.8342&60.6279&
 72&2s$^2$2p$^5$4f\ $^3{\rm G}_3$\ (1/2,5/2)3&&75.6139&75.3975\\
 28&2s2p$^6$3s\ $^3{\rm S}_1$\ (1/2,1/2)1&&63.3223&63.2125&
 73&2s$^2$2p$^5$4f\ $^3{\rm G}_4$\ (1/2,7/2)4&&75.6249&75.4057\\
 29&2s2p$^6$3s\ $^1{\rm S}_0$\ (1/2,1/2)0&&63.7906&63.6986&
 74&2s$^2$2p$^5$4f\ $^3{\rm F}_3$\ (1/2,7/2)3&&75.6331&75.4147\\
 30&2s2p$^6$3p\ $^3{\rm P}_0^\ro$\ (1/2,3/2)$^\ro$0&&65.7252&65.6346&
 75&2s$^2$2p$^5$4f\ $^1{\rm D}_2$\ (1/2,5/2)2&&75.6316&75.4155\\
 31&2s2p$^6$3p\ $^3{\rm P}_1^\ro$\ (1/2,1/2)$^\ro$1&65.601&65.7608&65.6676&
 76&2s2p$^6$4s\ $^3{\rm S}_1$\ (1/2,1/2)1&&81.6960&81.5889\\ %No. 114
 32&2s2p$^6$3p\ $^3{\rm P}_2^\ro$\ (1/2,3/2)$^\ro$2&&65.9265&65.8380&
 77&2s2p$^6$4s\ $^1{\rm S}_0$\ (1/2,1/2)0&&81.8528&81.7414\\  %No. 119
 33&2s2p$^6$3p\ $^1{\rm P}_1^\ro$\ (1/2,3/2)$^\ro$1&65.923&66.0704&65.9782&
 78&2s2p$^6$4p\ $^3{\rm P}_0^\ro$\ (1/2,1/2)$^\ro$0&&82.6666&82.5483\\ %No. 128
 34&2s2p$^6$3d\ $^3{\rm D}_1$\ (1/2,3/2)1&&69.0140&68.9221&
 79&2s2p$^6$4p\ $^3{\rm P}_1^\ro$\ (1/2,1/2)$^\ro$1&82.52&82.6786&82.5594\\
 35&2s2p$^6$3d\ $^3{\rm D}_2$\ (1/2,3/2)2&&69.0332&68.9323&
 80&2s2p$^6$4p\ $^3{\rm P}_2^\ro$\ (1/2,3/2)$^\ro$2&&82.7438&82.6281\\
 36&2s2p$^6$3d\ $^3{\rm D}_3$\ (1/2,5/2)3&& 69.0657&68.9518&
 81&2s2p$^6$4p\ $^1{\rm P}_1^\ro$\ (1/2,3/2)$^\ro$1&82.67&82.7908&82.6724\\
 37&2s2p$^6$3d\ $^1{\rm D}_2$\ (1/2,5/2)2&69.282$\dag$&69.4386&69.3247&%69.292 from Apj 502
 82&2s2p$^6$4d\ $^3{\rm D}_1$\ (1/2,3/2)1&&83.9021&83.7838\\  %(1998)1015
 38&2s$^2$2p$^5$4s\ $^3{\rm P}_2^\ro$\ (3/2,1/2)$^\ro$2&&71.8688&71.6517&
 83&2s2p$^6$4d\ $^3{\rm D}_2$\ (1/2,3/2)2&&83.9097&83.7886\\
 39&2s$^2$2p$^5$4s\ $^1{\rm P}_1^\ro$\ (3/2,1/2)$^\ro$1&71.860&71.9139&71.6983&
 84&2s2p$^6$4d\ $^3{\rm D}_3$\ (1/2,5/2)3&&83.9229&83.7976\\
 40&2s$^2$2p$^5$4p\ $^3{\rm S}_1$\ (3/2,1/2)1&&72.7909&72.5803&
 85&2s2p$^6$4d\ $^1{\rm D}_2$\ (1/2,5/2)2&&84.0529&83.9258\\
 41&2s$^2$2p$^5$4s\ $^3{\rm P}_0^\ro$\ (1/2,1/2)$^\ro$0&&72.7944&72.5824&
 86&2s2p$^6$4f\ $^3{\rm F}_2^\ro$\ (1/2,5/2)$^\ro$2&&84.4775&84.3462\\
 42&2s$^2$2p$^5$4s\ $^3{\rm P}_1^\ro$\ (1/2,1/2)$^\ro$1&72.74&72.8181&72.6062&
 87&2s2p$^6$4f\ $^3{\rm F}_3^\ro$\ (1/2,5/2)$^\ro$3&&84.4799&84.3481\\
 43&2s$^2$2p$^5$4p\ $^3{\rm D}_2$\ (3/2,1/2)2&&72.8545&72.6449&
 88&2s2p$^6$4f\ $^3{\rm F}_4^\ro$\ (1/2,7/2)$^\ro$4&&84.4858&84.3522\\
 44&2s$^2$2p$^5$4p\ $^3{\rm D}_3$\ (3/2,3/2)3&&72.8992&72.6948&
 89&2s2p$^6$4f\ $^1{\rm F}_3^\ro$\ (1/2,7/2)$^\ro$3&&84.4962&84.3621\\
 45&2s$^2$2p$^5$4p\ $^1{\rm P}_1$\ (3/2,3/2)1&&72.9287&72.7243\\
\noalign{\smallskip}
\hline
\end{tabular}
\end{center}
\end{table}
}

\newpage

\begin{table}  
\caption{Comparisons of weighted oscillator strengths (gf) for selected dipole allowded
transitions for  Fe~{\scriptsize XVII} in length-form (L) and velocity-form (V). `SS'
indicates present {\scriptsize SUPERSTRUCTURE} calculations;
`MCDF' results are obtained with {\scriptsize GRASP}; `NIST'
values are the recommended values from NIST website; `CIV3' are from \cite{hi93}
and `MBPT' are from \cite{sa02}; the quantity $a$e$b$ stands for $a\times10^b$.} 
\begin{center}
\begin{tabular} {cclllllll}
\hline
&&\multicolumn{2}{c}{SS}&\multicolumn{2}{c}{MCDF}&NIST&CIV3&MBPT\\
$i$&$j$&L&V&L&V&&L&L\\
\hline
3&1&0.124&0.112&0.125&0.121&0.122&0.122&0.127\\
5&1&0.102&0.100&0.106&0.101&0.105&0.102&0.105\\
17&1&8.70e-3&8.11e-3&1.01e-2&9.35e-3&9.7e-3&1.06e-2&9.85e-3\\
23&1&0.590&0.558&0.628&0.590&0.63&0.604&0.661\\
27&1&2.571&2.450&2.503&2.357&2.31&2.353&2.279\\
31&1&3.15e-2&3.20e-2&3.57e-2&3.59e-2&2.95e-2&3.33e-2&3.93e-2\\
33&1&0.280&0.296&0.282&0.283&0.282&0.278&0.278\\
39&1&2.47e-2&1.84e-2&2.54e-2&2.14e-2&2.54e-2&&\\
42&1&1.49e-2&1.37e-2&1.79e-2&1.47e-2&2.11e-2&&\\
52&1&3.57e-3&3.16e-3&3.88e-3&3.52e-3&3.67e-3&&\\
59&1&0.408&0.370&0.390&0.357&0.401&&\\
71&1&0.495&0.461&0.463&0.425&0.531&&\\
79&1&1.39e-2&1.39e-2&1.31e-2&1.27e-2&1.58e-2&&\\
81&1&0.100&0.104&8.88e-2&8.50e-2&0.115&&\\
6&2&0.252&0.232&0.256&0.242&0.248&0.250&\\
7&2&0.260&0.261&0.260&0.271&&0.253&\\
8&2&0.812&0.824&0.825&0.990&0.819&0.808&\\
7&3&0.284&0.341&0.287&0.362&&0.284&\\
9&3&0.322&0.303&0.327&0.350&&0.318&\\
10&3&0.281&0.244&0.283&0.295&&0.273&\\
11&3&0.102&7.43e-2&0.102&8.70e-2&&0.102&\\
15&3&7.93e-2&3.90e-2&7.98e-2&6.18e-2&&7.41e-2&\\
14&5&0.589&0.562&0.595&0.671&&0.581&\\
15&5&0.133&7.82e-2&0.134&0.104&&0.130&\\
\hline
\end{tabular}
\end{center}
\end{table}

%\newpage

\begin{table}  
\caption{Selected transition probabilities $A\cdot$s of Fe~{\scriptsize XVII}:
the first and second entries for each E1 transition are the respective values of
$A$ in the length-form and velocity-form; the quantity $a$e$b$ stands for $a\times10^b$.} 
\begin{center}
\begin{tabular} {cccc|cccc}
\hline
  $i$ &  $j$  &type& $A_{ij}$ &
  $i$ &  $j$  &type& $A_{ij}$ \\
\hline
3&1&E1&9.63e11& 5&1&E1&8.38e11 \\
&&&9.24e11& &&&8.02e11 \\
17&1&E1&9.42e10&23&1&E1&6.01e12 \\
&&&8.73e10& &&&5.65e12 \\
27&1&E1&2.47e13&33&1&E1&3.29e12 \\
&&&2.32e13& &&&3.30e12 \\
7&1&E2&5.24e8&6&1&M1&1.80e5\\
10&1&E2&5.63e8&9&1&M1&6.81e3\\
14&1&E2&6.77e8&12&1&M1&4.24e3\\
35&1&E2&1.86e7&13&1&M1&2.03e5\\
37&1&E2&1.09e10&28&1&M1&1.93e4\\
85&1&E2&3.00e9&34&1&M1&2.10e3\\
20&1&E3&2.83e5&2&1&M2&2.25e5\\
22&1&E3&3.52e5&18&1&M2&6.16e6\\
26&1&E3&4.00e5&21&1&M2&1.13e6\\
56&1&E3&3.87e4&24&1&M2&4.47e5\\
87&1&E3&1.23e5&25&1&M2&2.73e5\\
89&1&E3&3.36e6&32&1&M2&8.44e5\\
\hline
\end{tabular}
\end{center}
\end{table}

\subsubsection{Energies}

The calculated energies in Table 1 from \sss and {\small GRASP} are compared
with observed values wherever available. 
Table 1 also provides the key to the level indices for transitions in
tabulating transition probabilities, collision strengths, and
Maxwellian averaged collision strengths.              
The comparison with experimental values shows agreement
to within 1\% or better. The target representation is thus generally
better than in most EIE calculations.

\subsubsection{Transition probabilities}

We report here briefly our calcultions of transiton probabilities. Full
results will be reported elsewhere due to the space limitation.
Table 2 presents weighted oscillator strengths for a number of E1 transitions, whose accuracy 
indicates that of the target eigenfunctions.
Table 3 gives a few selected transition probabilities 
of dipole-allowed E1
transitions, forbidden magnetic dipole M1 and electric quadrupole E2 
transitions, and some unexpectedly strong E3 and M2 transitions.
Only E2 and E3 in length-from are tabulated in Tanle 3.
In intermediate-coupling, or in jj-coupling, intercombination transitions 
are a special type of E1 transitions.
The excellent agreement of the length-form (the first entry) and the velocity-form (the
second entry) A-values for each E1 transitions selected in Table 2 is a further prove that
we have obtained a good target for EIE calculations.
In Table 3, all selected E3 and M2 transitions are of order $10^4$ s$^{-1}$
or more. They may have some influence on the modeling of line emissions. In particular,
the M2 line from level 2-1 has long been observed as a prominent line
in astronomy and in laboratory photoionized plasmas,
The population of level 2 is fed up by
the cascade effects from 2p$^5$3s, 2p$^5$3p, and 2p$^5$3d and other higher configurations.
Accurate M2 transition probability is key to model this line.
This line also has an important plasma diagnostics potential.

\subsection{Electron impact excitation of \ \Fe17}

\subsubsection{37CC collision strengths}

In order to understand the dependence of resonances on the n-complexs,
we first calculate a set of 37CC collision strengths. 
Also, there are many previous 37-level (R)DW calculations so
it is more straightforward to compare previous results with the present 
37CC calculations.

Fig.~2 shows extensive resonances in several transitions. Fig.~2(a) is
the collision strengths for the
first resonant excitation $\Omega(^1S_0$-$^3$$P_2)$, a magnetic quadrupole (M2)
transition.
The dominant role of resonances is clear, when compared with the background collision
strengths and in particular compared to the earlier (R)DW calculations, which 
correspond only to the non-resonant background. 
Fig.~2(b) is for transition from the ground state to the 15th energy level 2p$^5$3p $^1$S$_0$.
This transition is a monopole transition which is very important in 
soft {\small X}-ray lasing studies.
The green dots are RDW calculations in \cite{zh89}, and the green squares are 
our RDW calculations
\cite{cgx}. Both RDW calculations agree very well with the background collision
strengths of the BPRM results.

\subsubsection{Comparison with non-resonant approximations:}

 We discuss in particular the RDW calculations as
representative of a long list of previous
calculations in literature using the
distorted-wave (DW) method, the Coulomb-Born (CB) method, or
their variants.
Although the (R)DW values are in very good agreement with the background values in the
present calculations, the dense resonance structures are not accounted for.
However, the relativistic effects and correlation effects in the target are
considered in the 
for in RDW calculations. The RDW method may be thought of as a 
2CC calculation including initial and scattered channels, without
the remaining channel coupling and resonance effects which, as we
demonstrate in this work, are not only significant but may dominate
certain transitions.
Since DW results are still extensively used
in spectral modeling, care should be exercised to take these 
limitations into account. 

The one previous non-relativistic
R-matrix calculation also agrees well with the present background values in the
high-energy region above the highest threshold in the 27-level target. 
Collision strengths are reported 
at a few energies in Ref.~\cite{mo97} in this region. It is clear that
since all calculations are above the highest threshold in the CC
expansion, there are no resonance structures included in the previous 
R-matrix work.

It should be noted that although for the 3 transitions 3C, 3D and 3E 
discussed in detail here,
the RDW results are in excellent agreement with the background values of the
89CC BPRM calculations,
there are very large discrepancies for some other transitions even in the background values due to
broad diffuse resonance or background enhancement. 
The differences can be
from a factor of two to up to an order of magnitude, depending on the
strength of the transition.
(see Figs.~12 for further discussions).

\subsubsection{Partial wave expansion: the 89CC collision strengths}

Before presenting the collision strengths from the 89CC calculations, 
it is instructive to point out the issues of
the convergence of partial wave expansion addressed in detail in our
study. We used the partial wave
collision strength $\Omega_J$, and the partial sum of PW collision strengths
$\sum_{1/2}^J\Omega_J$, as shown in Figs.~3,4 and 5.
All partial cross sections are from the 89CC BPRM calculations unless otherwise
indicated.

In Figs.~3(a)-(d), $\Omega_J$ and $\sum_{1/2}^J\Omega_J$ for the
transitions 3C and 3G are plotted
as a function of 2J. For each transition, $\Omega_J$ and $\sum_{1/2}^J\Omega_J$
are calculated at four incident electron energies: 100 Ry (black curve), 
200 Ry (green), 300 Ry (blue), and 400 Ry (Red). The filled circles in 
Figs.~3(b) and (d) are the corresponding total collision strengths that partial sum of PW collision strength
$\sum_{1/2}^J\Omega_J$ should converge onto. It is noted that there
is a peak and a minimum for all curves of $\Omega_J$ for the 3C
transition, and there are also some irregular
features around $7<2J<17$. The higher partial waves
($2J\ge53$) not calculated in this work all fall in the 
tail part of the $\Omega_J$ curves. This is necessary to ensure 
accurate CBe top-up (see Sec.~3, also discussed
later). It is found that for the 3C
convergence is achieved; the largest difference is $<5$\%
at $E_i=400$ Ry.

Figs.~3(b)and (d) show the partial wave convergence for the 3F transition. 
Again, there is one minimum for
most curves of $\Omega_J$; however, there are two peaks
and two minima $E_i=100$ Ry. Furthermore, 
the 3F curves are much more irregular than the 3C, as
seen from Figs.~3. When we check the convergence from Figs.~3(b) and (d), we 
find that at low $E_i$ convergence has been achieved. 
For higher $E_i$ however, complete convergence
is not obtained in the CC calculations. But the positions of the last peak 
for all the $\Omega_J$ curves are well below the highest partial wave 
(2J=51) employed in the
present 89CC calculations, so the CBe top-up approach should again be 
very accurate and we have
convergent collision strengths for these types of transitions.

The transitions shown in Figs.~3 are from the ground level
to the \n = 3 levels. In Figs.~4(a)-(d), we discuss further
a particular transition, 2-44 (indices in Table 1,) 
between excited levels from \n = 3 to \n = 4. Similar features as
the curves of $\Omega_J$ above are also found in Fig.~4(a). 
Noticeable differences are: (i) the positions of the peaks are at much 
higher J values (close to the highest
J used in CC calculations), in particular for high $E_i$ energies; 
(ii) the values of $\Omega_J$ at 2J = 51 are still a big fraction of the 
peak values. This feature implies that the CC results have certainly 
not converged, as can also be 
proved from the partial sum of $\sum_{1/2}^J\Omega_J$ in Fig.~4(b).
In fact, at $E_i$=400 Ry, $\sum_{1/2}^J\Omega_J$ is only $\sim$50\% of the 
converged value for the
transition 2-44. The main reason for this difference,
compared to Figs.~3, is that the transition energies for \n = 3-4
are much smaller than for the \n=2-3. Fortunately, the
positions of all highest peaks for all $\Omega_J$ curves are located below 
(though close to)
2J=51, so the CBe top-up approach should be accurate. 
To reconfirm this,
we did a RDW calculation for the transition 2-44 including partial
waves up to 80. The results are plotted in Figs.~4(c) and (d). We have a
slightly different notation for the RDW results. In Fig.~4(c),
the partial wave $\Omega_\ell$ is plotted as a function of the 
orbital angular momentum $\ell$ of
scattered free electron; while in Fig.~4(d), the partial sum of
$\Omega_\ell$ is plotted as a function $\ell$. 
However, according to the collision
theory of partial wave expansion, we arrive at an identical conclusion. 
As seen from
Fig.~4(c), convergent results have been obtained with partial waves
up to $\ell$=80 for the transition 2-44.

Partial wave analysis shown in Figs.~3-4 are for dipole allowed transitions.
It is expected that the convergence for E2 transitions would also be 
very slow. This can
be seen from the two E2 transitions, 2-51 and 6-60, 
shown in Figs.~5(a)-(d). Similar features to
Figs.~4 are found, and the same conclusion can be drawn.
In addition, we used the Burgess-Tully method \cite{bu92}
to study the convergence of the E2 cross sections
in Figs.~6, where the reduced collision strengths $\Omega_r$ 
are plotted as a function of reduced electron
energy $E_r=E_i/(E_i+c)$ (c=e is a constant, $E_r\in$ [0,1]), for the
E2 transitions 2-51 and 6-60.
The red filled circle corresponds to infinite incident energy;
our RDW results are also shown for comparision as blue squares.
From this figure we conclude that the BPRM calculations of E2 
transitions 2-51 and 6-60 are well converged for  E$_i<400$ Ry
(E$_r<$0.7).

\subsubsection{89CC Collision strengths}

 The 89CC calculations reveal the presence of
resonances due to the \n = 4 levels that appear not only above the \n =
3 thresholds, but also below those energies. This in particular affects
the comparison of theoretical results with the recent EBIT measurements
of relative line ratios, as discussed in \cite{ch02}.

Below, we discuss several aspects of the large 89CC calculations.

{\it A. Comparison of 37CC and 89CC results:}

The differences between the two sets of calculations are illustrated
for the three transitions 3C, 3D, and 3E in
Figs.~7-9, respectively, that correpsond to important 
{\small X}-ray lines observed in laboratory and astrophysical plasmas.
The RDW calculations and non-relativistic
R-matrix calculation \cite{mo97} without resonances
are also shown for comparison.

The 3C is the strongest resonance
transition from the ground state in the emission or absorption spectra
of \Fe 17. Figs.~7(a) and 7(b) show the 37CC and the 89CC collision strength
respectively.
The green open square is the non-relativistic DW calculation \cite{bh92},
and the blue triangles are previous 27CC R-matrix calculations \cite{mo97,gu00}.
In Fig.~7(c), the 37CC collision strengths (red) is overlapped with the 89CC
results in order to show the difference between them.
Three main differences are noticed:
(i) Resonances in the range of 69 Ry$<$E$_i<$75 Ry show up in 89CC calculations due to
the inclusion of targets states in configurations 2p$^54\ell(\ell=$s,p,d,f).
Practical implication of this point are discussed later.
(ii) Denser resonances appear in the 89CC calculations for E$_i<$69 Ry. This
is also important in comparing the relative line intensity measured in EBIT experiments,
and in obtaining correct effective collision strengths under different
electron velocity distributions (e.g. gaussian, maxwellian, numerical).
(iii) The background collision strengths are lower in the 89CC calculations. 
This is due to the effect of channel coupling.
Redistribution of electron fluxes at a larger number of thresholds, as
opposed to a smaller number, reduces the background.
Although there is only a few percent difference in the 3C background 
collision strengths,
it is the strongest transition which is a factor of a few to
orders of magnitude larger than other transitions. Therefore
the redistribution of the 3C electron collision flux results in a significant,
10\% or more, change for other transitions.

The green and red dashed-lines shown in the plot are the 
numerically averaged (NA) and the gaussian averaged (GA with a 30 eV FWHM) 
collision strengths used to calculate line intensity ratios discussed
in Sec.~6.1. The NA and GA values can also be used to assess resonance 
enhancements due to the 89CC
calculations, from Figs.~7(a) and (b).

Figs.~8 and 9 provide details of resonance enhancements for the
intercombination transitions 3D and 3E, respectively.
The RDW calculations \cite{zh89} and R-matrix calculations \cite{mo97,gu00} are 
also shown. Several features may be noted.
(i) Resonance
enhancements in 3D and 3E in the 89CC calculations are much
more pronounced than for the 3C.
Clearly, the \n = 4 levels add considerably more resonances, both 
above and {\it below} the \n = 3 complex.
Again, similar to the case for 3C in Figs.~7,
these effects are more readily discernible by comparing the NA and GA 
values from the respective 37CC and 89CC calculations 
(relative to their respective backgrounds).
(ii) While both the 3D and 3E are intercombination
transitions, with denser resonances than the 3C, the energy behavior of
all three transitions is different.
The 3D collision strength increases with E$_i$,
with basically the same energy-dependence as the dipole allowed 3C, but the
3E decreases with increasing E$_i$ over a very broad energy range,
typical of a forbidden transition.
(iii) The physically different energy behaviour is also related to the 
appropriate coupling scheme that should be used to describe these three 
transitions in each case. Whereas the dipole allowed transition 3C 
can be treated in either LS or jj-coupling scheme,
the 3D is more appropriatly considered in a relativistic jj-coupling scheme.
However, a non-relativistic LS-coupling scheme may describe the behaviour
of the 3E (spin) forbidden transition. These effects demonstrate that
the atomic structure and
collision dynamics in intermediate Z and medium-to-highly
charged many-electron atomic species such as \Fe17 are rather complicated,
since they lie in a transient region where relativisitc effects begin to
manifest themselves. 

 Finally, we note that the dominant role of the \n = 4 resonances
seen in the 89CC results in Figs.~7-9, is due to  resonant configurations
$2s^22p^53\ell4 \ell'$
that manifest themselves from $\sim$47 Ry, considerably {\it
below} the excitation thresholds of the 2p \lr 3d lines 3C, 3D, and 3E
at $\sim$60 Ry. Therefore there are relatively fewer $n = 3$
resonances, and the $n = 4$ resonances greatly influence the
near-threshold behavior of these cross sections. In the region
75 - 84.5 Ry the resonances are
due to thresholds corresponding to two-electron-excitation
configurations $2s2p^64\ell$ with weakly coupled channels and
resonances, and should be affected by the higher $n > 4$ resonances.

{\it B. Wavefuncion expansion coefficients:}

 The $E_i$-dependence of 3C, 3D, and 3E can also be
quantitatively addressed from the eigen function expansion of the upper levels
in the \Fe17 target.  We have:
upper level (27) of 3C: $0.7857|27\rangle+0.1753|23\rangle+0.0305|17\rangle$;\ 
upper level (23) of 3D: $0.7479|23\rangle+0.2010|27\rangle+0.0491|17\rangle$;\ 
upper level (17) of 3E: $0.9150|17\rangle+0.0767|23\rangle+0.0030|27\rangle$.\ 
The transitions probabilities for 3C, 3D and 3E are 2.47e13, 6.01e12, and 9.42e10,
respectively. Level 17 is nearly described in pure LS-coupling from its
eigen function expansion, while levels 23 and 27 begin to considerably 
depart from pure LS-coupling to intermediate-coupling scheme (or jj-coupling).
It is the strong coupling of level 23 with level 27 that makes the 3D 
line behave as a dipole-allowed transition, as evident from the mixing
coefficients. In LS-coupling, the 3E is 
a spin-forbidden transition, so its collision strengths decreases 
with $E_i$.

{\it C. Other excited E1 transitions:}

In addition to the dipole allowed E1 transitions (including 
intercombination E1 transitions) discussed above,
two other interesting E1 transitions 1-5 (3F) and 1-33 (3A) are shown in 
Figs.~10.
The upper level of the 3F (as the 3G in Fig.~3) is strongly mixed,
therefore LS coupling designations are not appropriate, as discussed in Sec.~1.
Instead, we need to use jj-coupling notation. 
Because of the strong coupling in the target,
the two transitions show similar resonance structures. 
Both exhibit the allowed transition
behavior, although the 3F is in fact an intercombination transition. 
The RDW results and previous R-matrix results are also shown for comparison
(Fig.~10a), with good agreement for the background.

The upper level of the transition 3A (Fig.~10b) is from a configuration 
with a 2s-hole, therefore the strong channel coupling effects 
with the 2s2p$^5$4$\ell$ ($\ell$=s,p,d,f) configurations
cause the full range of pronounced resonances up to the highest threshold of the
\n=4 complex.

{\it D. Electric quadrupole (E2) and magnetic dipole (M1) transitions:}

In Fig.~11, collision strengths for six forbidden transitions 
(pure E2, pure M1, mixed E2 and M1) are displayed.
In Figs.~11(a) and (b) two pure E2-type transitions from the ground level, 1-7 and 1-37,
are shown. Resonant features in 1-37 are similar to 1-33 since the upper 
level of 1-37 also has a 2s-hole (as discussed in subsection C above). 
In Fig.~11(a), one previous R-matrix value for 1-7
is {\it smaller} than the present background, and all RDW values, 
by more than a factor of two.
In Figs.~11(c) and (d), two pure M1-type transitions from the ground level,
 1-6 and 1-9, are shown. In Fig.~11(d), one previous R-matrix value for 1-9
is {\it larger} than the present background and the RDW values by more 
than a factor of two.
In Figs.~11(e) and (f), two mixed E2 and M1-type transitions
between excited levels 2-3 and 2-5, are shown. Although there is no 2s-hole 
in the upper levels 3 and 5, it is interesting that some prominent resonant 
features appear near the highest threshold of the \n=4 complex.

{\it E. Electric octupole (E3) and magnetic quadruple (M2) type transitions:}

 A particularly noteworthy finding of the present work is the potential
significance of the E3 transitions in spectral modeling of \Fe17
based on their relatively large A-values
(Table 3). In Fig.~12, four forbidden transitions (pure E3, pure M2, 
mixed E3 and M2, and mixed E1, E3 and M2 types) are given.
In Fig.~12(a), a pure M2-type transition from the ground level 1-2 is shown.
This transition is also presented in Fig.~2(a) from the 37CC calculations. 
One previous non-resonant R-matrix value
for 1-2 is {\it smaller} than the present background and all the RDW 
values by more than
a factor of two \cite{not1}. The RDW values agree with the present background in the 
high-energy region. However, in the low-energy region, 
the RDW values differ from the present background by about a factor of two.
In Fig.~12(b), a pure E3-type transition from the ground level 1-20 is shown.
All RDW results, and the previous non-resonant R-matrix values, are in good 
agreement with the present background.

A mixed M2 and E3-type transition between excited levels 3-36 is shown
in Fig.~12(c), and a mixed E1, M2 and E3-type transition between 
excited levels 2-35 is shown in Fig.~12(d). As the cases discussed earlier, 
the upper levels of 3-36 in Fig.~12(c), and 2-35 in
Fig.~12(d) have a 2s-hole leading to pronounced resonances in the
energy range up to the \n = 4 complex for both transitions. 
The RDW calculations are
off by more than a factor of two for both transitions; 
while the previous DW values
are differ by an order of magnitude, as shown in the figures.

{\it F. The Monopole transitions:}

Finally, we illustrate the monopole transitions of great interest in
laser excitations in \Fe17 since they are collisionally, and not
radiatively, excited. In Fig.~13, two monopole transitions are given.
In Fig.~13(a), the monopole transition from the ground level 1-15 is shown.
This transition is also presented in Fig.~2(b) from the 37CC calculations.
In Fig.~13(b), the monopole transition from the ground level 1-11 is shown.
The RDW values differ slightly but the previous non-resonant 27CC R-matrix values
differ considerably with the present background collision strengths.

\section{Rate coefficients and line intensity ratios}

Significant resonance enhancement of the collision strengths of forbidden and
intercombination transitions has been demonstrated in this work.
This directly enhances the rate coefficients which, in turn, affect the
computation of line intensities. While the calculation of rate
coefficients is a voluminous task, currently under way (to be reported another
paper), we present selected results from this study and apply them
to the analysis of previous experiments and observations.

\subsection{Comparisons with EBIT experiments}

 As described earlier, two sets of EBIT experiments have measured the
relative intensities of the 3C, 3D, and 3E transitions 
\cite{br98,la00}. The monoenergetic beams used in these experiments 
sample the effective cross sections averaged over the beam width.
Although the beam width is relatively large ($\ge$30 eV FWHM) compared to the
resonance widths, unlike astrophysical plasmas with typically Maxwellian
distribution over a wide range of electron velocities,
the monoenergetic EBIT experiments accurately probe the major 
atomic processes in line formation: electron impact excitation and 
radiative decay, and are not affected by other processes.
Line intensities measured in the EBIT experiments for 
3C, 3D and 3E transitions can be modeled as \cite{br98,not2}
\begin{eqnarray}
I_{k1}=\eta(\lambda_{k1})B_{k1}\langle\sigma_{k1}\cdot v\rangle n_eN_{\rm \Fe17}
\end{eqnarray}
where the upper state k is 27, 23 and 17 respectively for the three
transitions. Response function $\eta(\lambda_{k1})$ of the spectrometer 
is taken to be the same for all the three lines because the wavelengths of the three 
lines --- 15.01$\AA$, 15.26~$\AA$, 15.46~$\AA$ --- are close together. 
The branching ratios for all the radiative
decay routes are calculated to be 1.0, 1.0, 0.89 for 3C, 3D and 3E, 
respectively.
$\langle\sigma_{k1}v\rangle$ is the beam averaged rate coefficient;
$n_e$ and $N_{\rm \Fe17}$
are the electron density and the density of \Fe17 ions respectively.
One then obtains \cite{not2} 
\begin{eqnarray}
R1=3C/3D=\langle\sigma_{3C}\cdot v\rangle/\langle\sigma_{3D}\cdot v\rangle \\
R2=3E/3C=0.89\langle\sigma_{3E}\cdot v\rangle/\langle\sigma_{3C}\cdot v\rangle 
\end{eqnarray}
For monoenergetic electron beams employed in EBIT experiments, 
we use two ways to calculate
the rate coefficients using the collision strengths in
Figs.~7-9 with complex resonance structures. The first is a direct numerical
average (NA), and the second is a Gaussian average (GA) using a beam
width of 30 eV \cite{br98}. The averaged results are not
very sensitive either way, owing to the fairly large beam widths,
although there is a small variation in some regions. 
Both the NA and GA averaging results
are also shown in panels (a)\&(b) in Figs.~7-9 as red and green curves 
respectively. From the working assumptions outlined above,
the ratios of line intensities in EBIT experiments are in fact the
ratio of their averaged-collision strengths convolved over the electron 
beam width, as in Eqs.~(4) and (5). 
We note particularly that the observed up and down `oscillation' in the
line ratio R1 seen in the EBIT experiments \cite{br01}
is likely $not$ from the experimental error bars, but from the 
averaged-collision strengths with varying
resonance enhancements in different energy regions, and/or
different types of transitions.
Both, the collision strengths for the 3C, 3D, or 3E, and the
line ratios R1 and R2 as a function of electron energy, 
have additional complexities
because of fine-structure and coupling effects.
Therefore, constant or simple ratios of collision strengths
at one energy, extrapolated to the entire energy range, are not accurate.

\begin{table}
\caption{Comparison of present line ratios for R1=3C/3D and R2=3E/3C with EBIT
measurements}
\begin{center}
\begin{tabular} {ccccc}
\hline
&&$E_i$=0.85 keV&0.9 keV& 1.15 keV\\
\hline
&EBIT&2.77$\pm$0.19$^a$&2.94$\pm$0.18$^b$&(3.15$\pm$0.17,2.93$\pm$0.16)$^a$\\
R1=3C/3D&Theory$^c$:NA&2.80&3.16&3.05$^\dag$\\
&Theory$^c$:GA&2.95&3.27&3.10$^\dag$\\
&Other Theory&\multicolumn{3}{c}{3.78$^d$;\ \ \ \ \ \ \ \ 
4.28$^e$;\ \ \ \ \ \ \ \ 3.99$^f$\ \ \ \ \ \ \ \ }\\
\hline
&EBIT&&0.10$\pm$0.01$^b$&\\%0.09$\pm$0.01$^b$\\
R2=3E/3C&Theory$^c$:NA&0.11&0.085&0.07$^\dag$\\
&Theory$^c$:GA&0.11&0.083&0.07$^\dag$\\
&Other Theory&\multicolumn{3}{c}{0.04$^d$;\ \ \ \ \ \ \ \ 
0.05$^e$;\ \ \ \ \ \ \ \  0.05$^f$\ \ \ \ \ \ \ \ }\\
\hline
\end{tabular}
\end{center}
$^a$ EBIT experiments at LLNL \cite{br98};
$^b$ EBIT experiments at NIST \cite{la00};
$^c$ present theory with NA and GA;
$^d$ \cite{zh89};
$^e$ \cite{bh92};
$^f$ \cite{mo97};
$^\dag$present values with extrapolation of resonance enhancement from
\abi collision strengths from E $\leq$ 1.02 keV (see text).
\end{table}

Table 4 shows the line intensity ratios calculated from Eqs.~(4) and (5) and the procedures
discussed above. The agreement with two independent
EBIT experiments is excellent, to 10\% or within the
experimental error bars. In the transitions 3C, 3D, and 3E we find 
strong resonances in the 89CC calculations for all energies up to
75.0 Ry (1020 eV), corresponding to level 75 (Table 1), indicating that
channel couplings in this region are accounted for. However, even higher
thresholds (e.g. \n = 5) are likely to contribute to resonances above 
this threshold. In order to compare with 
EBIT experimental data at these higher energies, we extrapolate 
the resonance enhancement to the region
above 75.0 Ry. The predicted values are also given in Table 4. 
It should be mentioned that although the highest threshold in the 89CC
calculations is $\sim$84.5 Ry, more practically we define the highest 
strong `resonance' threshold to be at 75 Ry because the levels 76-89 
are from configurations $2s2p^64\ell~(\ell=s,p,d,f$) 
and the channel-couplings with transitions 3C, 3D, and 3E is therefore very
weak, as expected from Figs.~7-9 in the energy range $E_i>75$ Ry \cite{not3}.
The physical implication of these weak couplings 
is that the flux transfer between channels arising
from two-electron-excitation processes is very weak in the lines 3C, 3D, and 3E.

\subsection{Comparison with observations from stellar coronae}

Next, we carry out the line intensity calculation using a 89-level
collisional-radiative model (CRM). The rate coefficients are
obtained by averaging the collision strengths over a Maxwellian velocity
distribution prevalent in most astrophysical plasmas,
unlike the approximately monoenergetic velocity distribution in EBITs. 
Although our line ratios are in good agreement with EBIT experiments, 
other atomic processes, particularly cascades from higher levels, may be
significant in astrophysical environments.

Fig.~14 is an example of the 89-level CRM results with the present 
relativistic excitation rates for \Fe17, and their potential applications 
to {\small X}-ray astronomy in sources such as stellar coronae.
The detailed collision strengths for the {\small X}-ray line 3F (1-5),
an intercombination transition, is shown in Fig.~10(a), with 
huge resonance effects. The temperature dependence of 
3F/3C ratio (the {\small X}-ray line 3C is an allowed transition) 
is demonstrated in Fig.~14, and compared
with previous calculations (filled squares).
The electron density dependence is small; solid-line and dot-line correspond
to 10$^{13}$ and 10$^9$ cm$^{-3}$ respectively. The 4 open
circles with error bars are observed and experimental values.
At all temperatures T $< 10^7$ K the present line ratio departs considerably
from previous calculations, to more than a factor of 3
at about 10$^6$ K---a fact of considerable importance in
photoionized {\small X}-ray plasmas that have
temperatures of maximum abundance much lower than that in coronal
equilibrium T$_m \sim  4 \times 10^6$ K for \Fe17, as marked.

We also find that using our Maxwellian-averaged 
rates the line ratios are lower than some of the 
astronomical observation data, in particular in the
non-flaring active region of solar corona \cite{hu76,ch02}. 
Our theoretical calculations thus support
the conclusion that some other physical processes contribute to level population
kinematics, e.g. level-specific recombination-cascades
from Fe~{\small XVIII} to \Fe 17 levels, blending of satellite 
lines due to inner-shell excitation of Fe~{\small XVI}, 
and resonant scattering of the 3C line \cite{br98,la00}.
Investigation of these mechanisms is still underway, however 
the present accurate \abi theoretical calculations set denifite limits
on atomic/astrophysical models to estimate opacity effects,
column densities, emission measures, and other astrophysical quantities.

\section{Conclusion}

 The principal features of the present work are as follows.

 (I) The hitherto most detailed sets of 37CC and 89CC BPRM calculations 
show that the \n = 4 complex explicitly included in the latter calculation
has a considerable effect on the
collision strengths for \Fe 17 transitions. In particular, prominent
resonances appear in the 89CC calculations at energies above and below
the \n = 3 thresholds, and the effective collision strengths are
considerably enhanced relative to the smaller 37CC calculation.
The background collision strengths for some transitions
are also affected due to inter-channel coupling and consequent
re-distribution of flux among the larger number of channels in the 89CC case.

 (II) New calculations of atomic structure and transition probabilities
for all 89 levels have also been carried out, including E1,E2,E3 and M1
and M2 multipole transitions. Some results are 
presented, although the primary
focus is on electron excitation. Owing to the complexity of Fe~{\small XVII}
relativistic and correlation effects are equally important. We note that
neither pure LS-coupling nor pure
jj-coupling is appropriate for some transitions.
It is found that the M2 and E3 transition probabilities are sufficiently
large and may have non-negligible
effect on the intensities of some important Fe~{\small XVII} lines.
 
 (III) The calculated effective collision strengths are benchmarked 
against EBIT experimental data, and show
very good agreement to $\sim$10\%, or within experimental uncertainties.
Resonance enhancements in the intercombination lines 3D and 3E
is much larger than for the dipole-allowed line 3C, and is crucial to 
spectral formation of these important lines. The strong
dependence of line ratios 3C/3D and
3E/3C on electron beam energy is explained by the present results.

\ack

This work was partially supported by the National Science Foundation
and the NASA Astrophysical Theory Program. The computational work was
carried out at the Ohio Supercomputer Center, Ohio.

\section*{References} 

\def\amp{{Adv. At. Molec. Phys.}\ }
\def\apj{{ Astrophys. J.}\ }
\def\apjs{{Astrophys. J. Suppl.}\ }
\def\apjl{{Astrophys. J. (Lett.)}\ }
\def\aj{{Astron. J.}\ }
\def\aa{{Astron. Astrophys.}\ }
\def\aas{{Astron. Astrophys. Suppl.}\ }
\def\adndt{{At. Data Nucl. Data Tables}\ }
\def\cpc{{Comput. Phys. Commun.}\ }
\def\jqsrt{{J. Quant. Spectrosc. Radiat. Transf.}\ }
\def\jpb{{J. Phys. B}\ }
\def\pasp{{Pub. Astron. Soc. Pacific}\ }
\def\mn{{Mon. Not. R. Astron. Soc.}\ }
\def\pra{{Phys. Rev. A}\ }
\def\ps{{Phys. Scr.}\ }
\def\prl{{Phys. Rev. Lett.}\ }
\def\zpds{{Z. Phys. D Suppl.}\ }
\def\adndt{At. Data Nucl. Data Tables}

%\begin{references}
%\begin{harvard}  

%\input fig.tex

%\pagestyle{empty}
\newpage
\begin{figure}
\centering
\psfig{figure=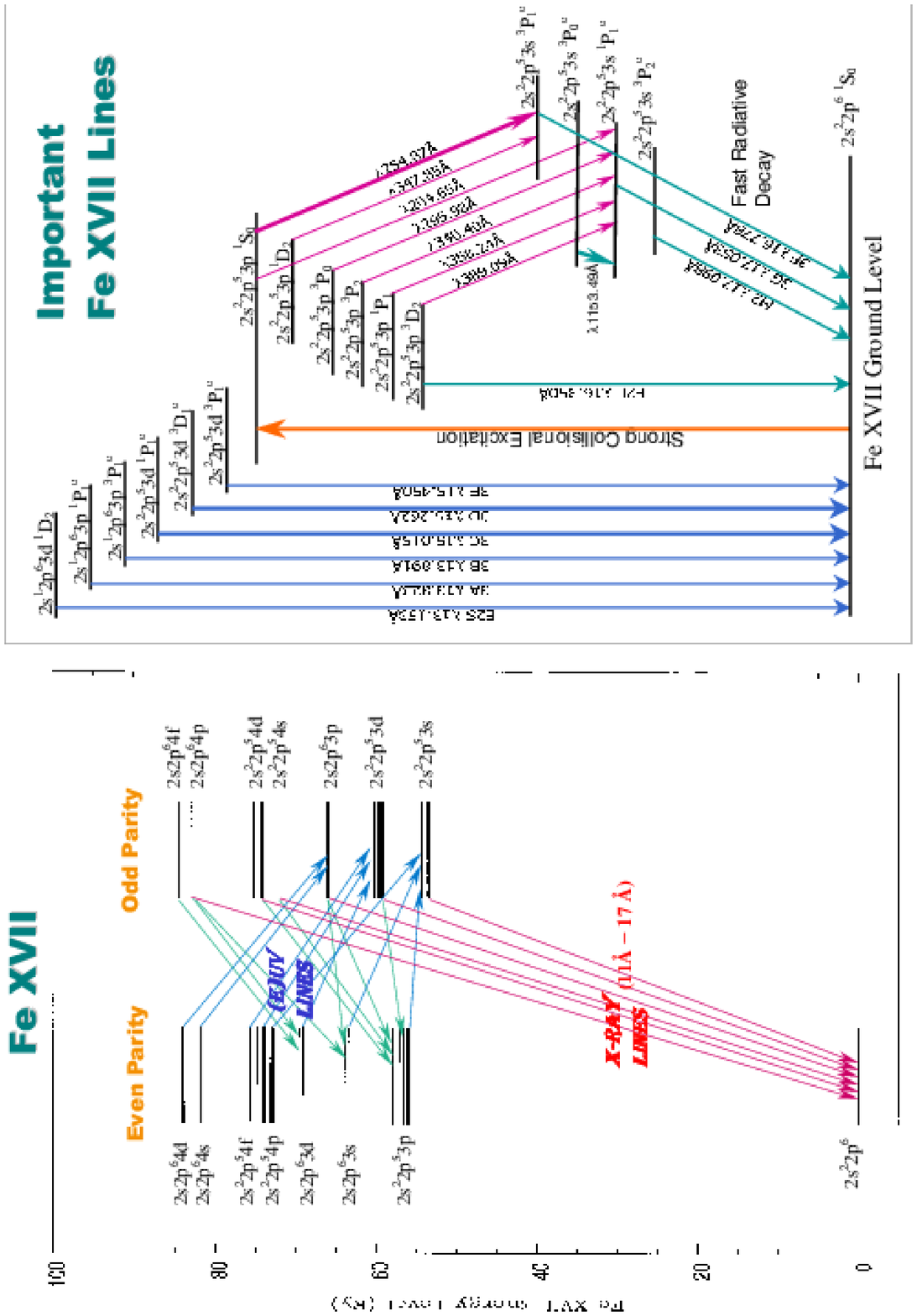,height=17.0cm,width=18.0cm}
\caption{Grotrian energy diagram and important \Fe17 {\small X/UV} lines.
(a) Schematic {\small X}-ray transitions to the ground level and {\small X/UV}-ray transitions
between excited levels; (b) Wavelengths and level designations for
some {\small X}-rays of considerable importance in astrophysics and for important
soft {\small X}-ray laser lines in laboratory plasmas are sketched.
}
\end{figure}

% Fig.2
\begin{figure}
\centering
\psfig{figure=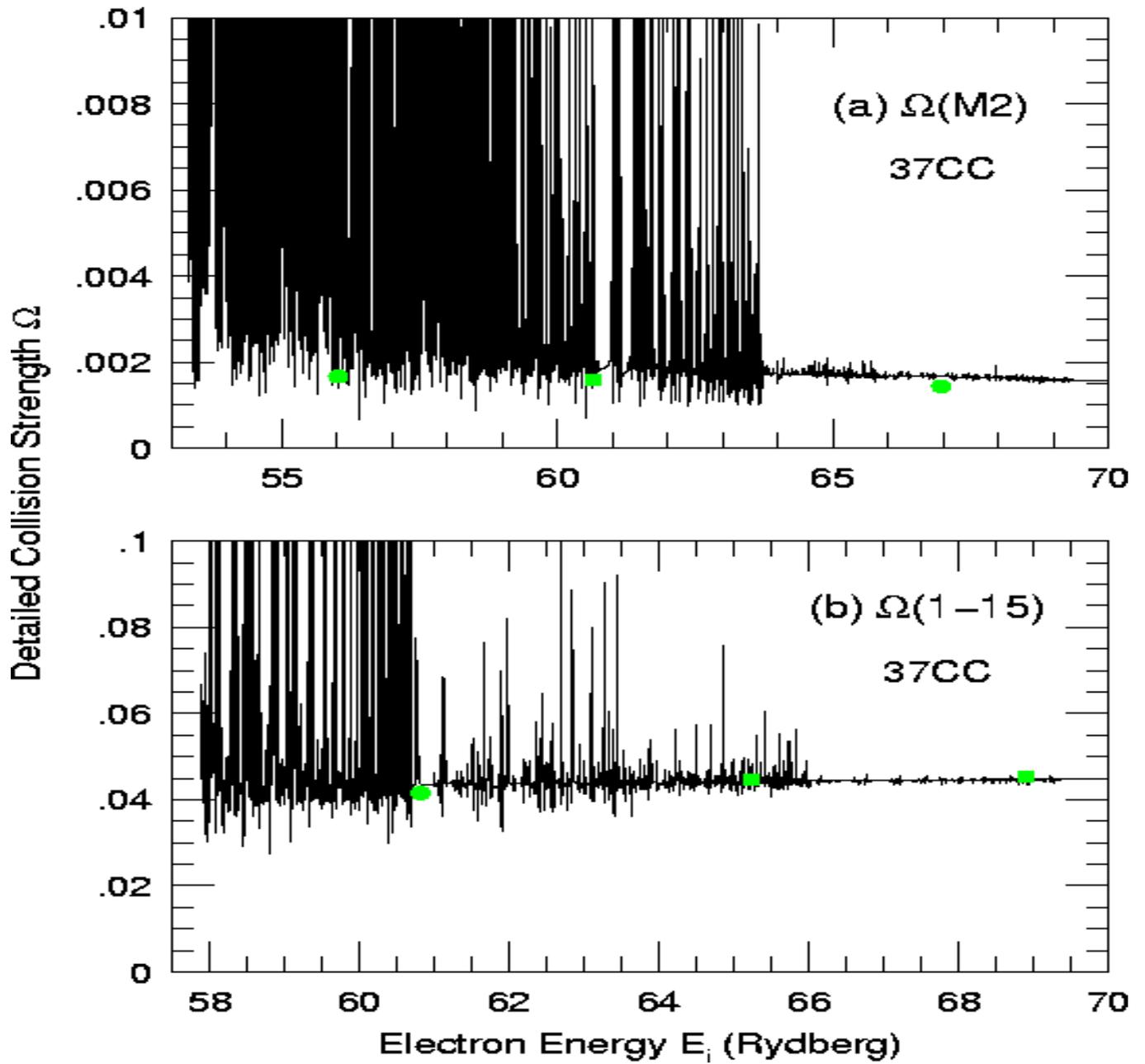,height=17.0cm,width=18.0cm}
\caption{Collision strength $\Omega$ from 37CC BPRM calculations
with detailed resonance structures as a function of incident electron energy $E_i$:
(a) magnetic quadrupole M2 transition 1-2; (b) monopole transition 1-15.
The green dots and sqaures are RDW values in \cite{zh89} and in \cite{cgx}, respectively.
}
\end{figure}

\begin{figure}
\centering
\psfig{figure=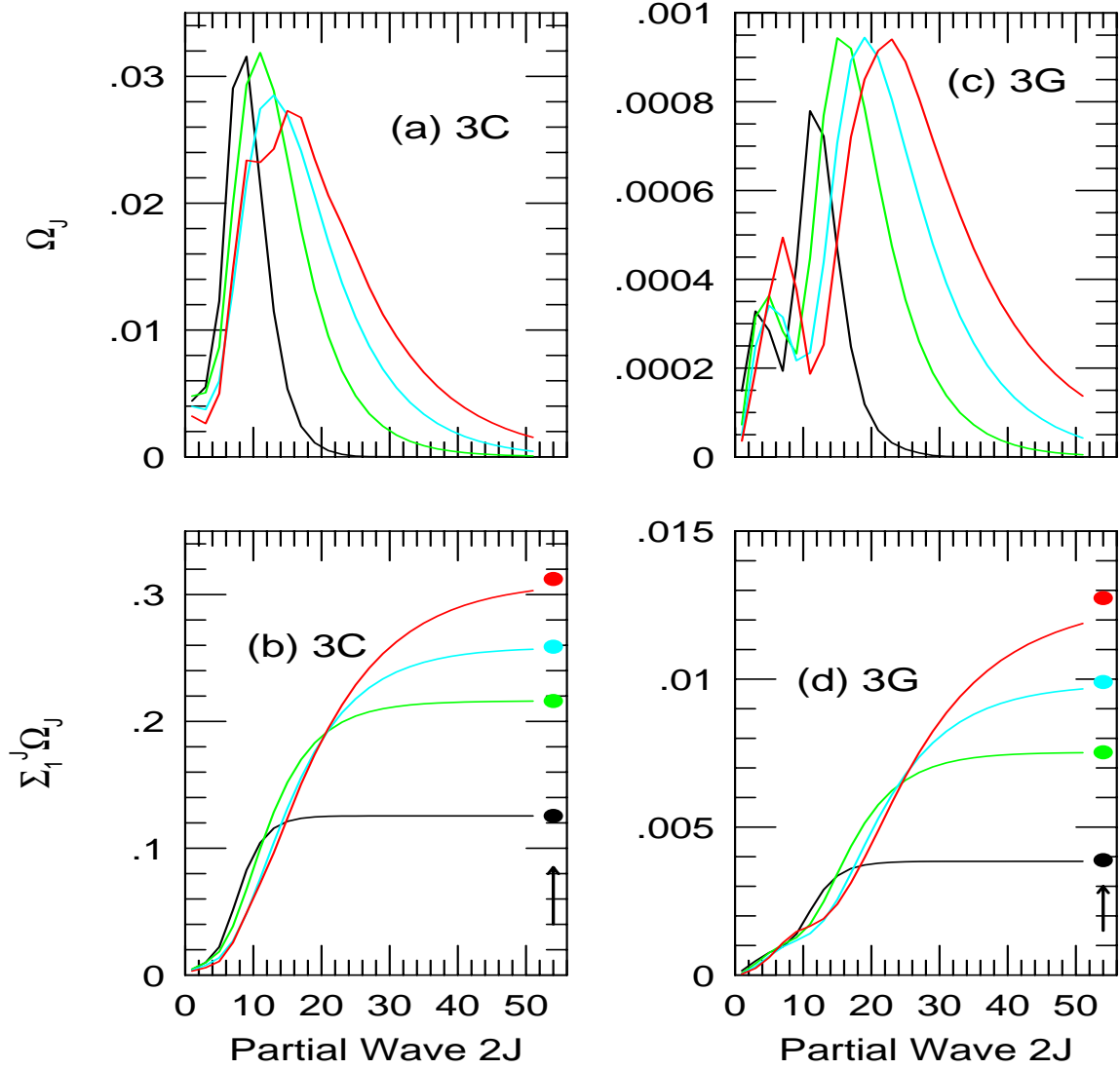,height=17.0cm,width=18.0cm}
\caption{Partial wave (PW)
collision strength $\Omega_J$ and partial sum of PW collision strength
$\sum_{1/2}^J\Omega_J$ for E1 transitions. (a) and (b): transition 3C(1-27);
(c) and (d): 3G(1-3). For each transition, $\Omega_J$ and $\sum_{1/2}^J\Omega_J$
are calculated at four $E_i$: 100 Ry (black curve), 200 Ry (green),
300 Ry (blue), and 400 Ry (Red). The filled circles in (b) and (d) are the
corresponding total collision strengths that partial sum of PW collision strength
$\sum_{1/2}^J\Omega_J$ should converge onto.
}
\end{figure}

\begin{figure}
\centering
\psfig{figure=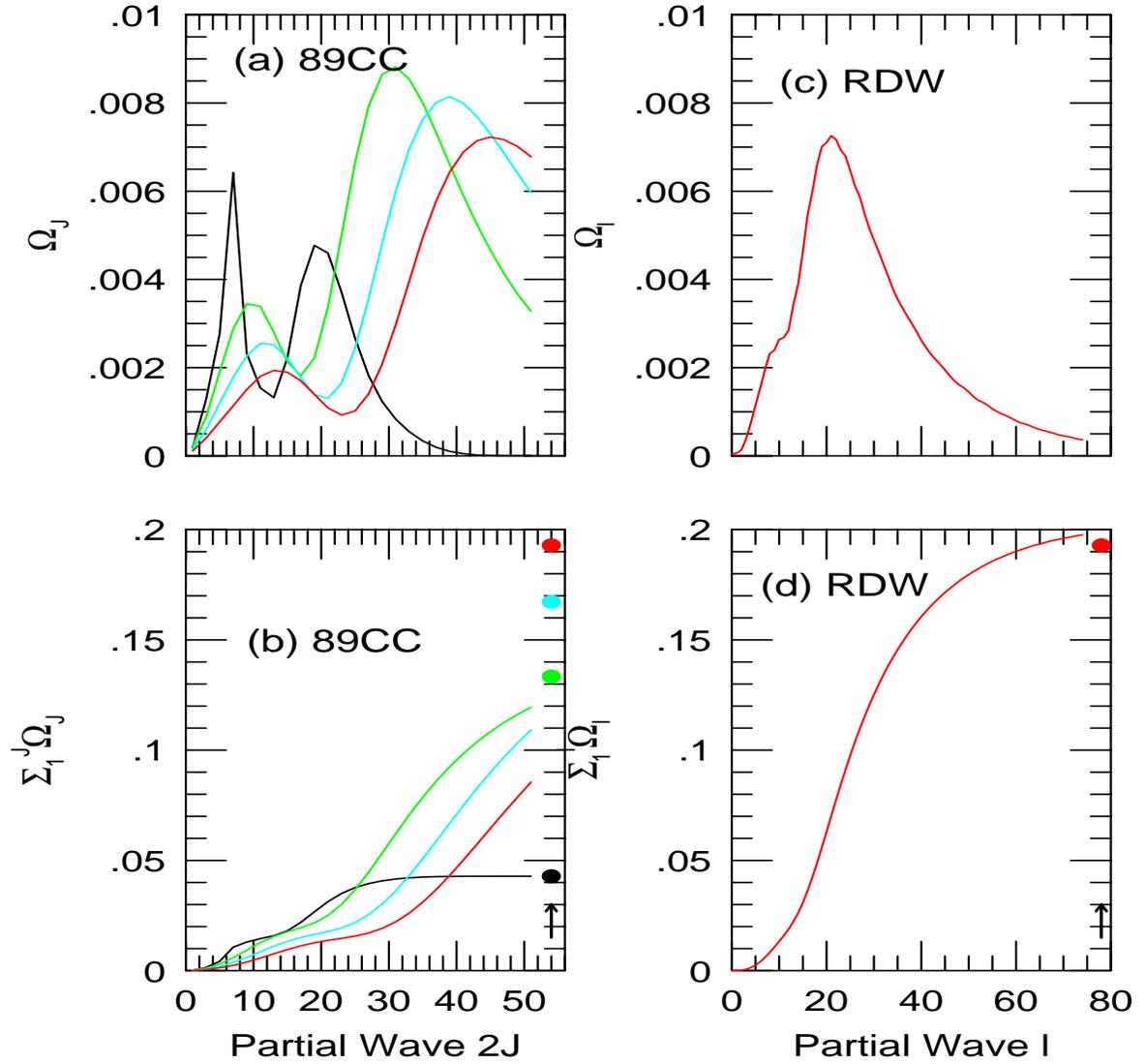,height=17.0cm,width=18.0cm}
\caption{Partial wave (PW)
collision strength $\Omega_J$ and partial sum of PW collision strength
$\sum_{1/2}^J\Omega_J$ for E1 transition between excited levels 2-44.
(a) and (b): from 89CC;
(c) and (d): $\Omega_\ell$ and $\sum_\ell\Omega_\ell$ from RDW \cite{cgx}.
Other symbols are the same as Figs.~3.
}
\end{figure}

\begin{figure}
\centering
\psfig{figure=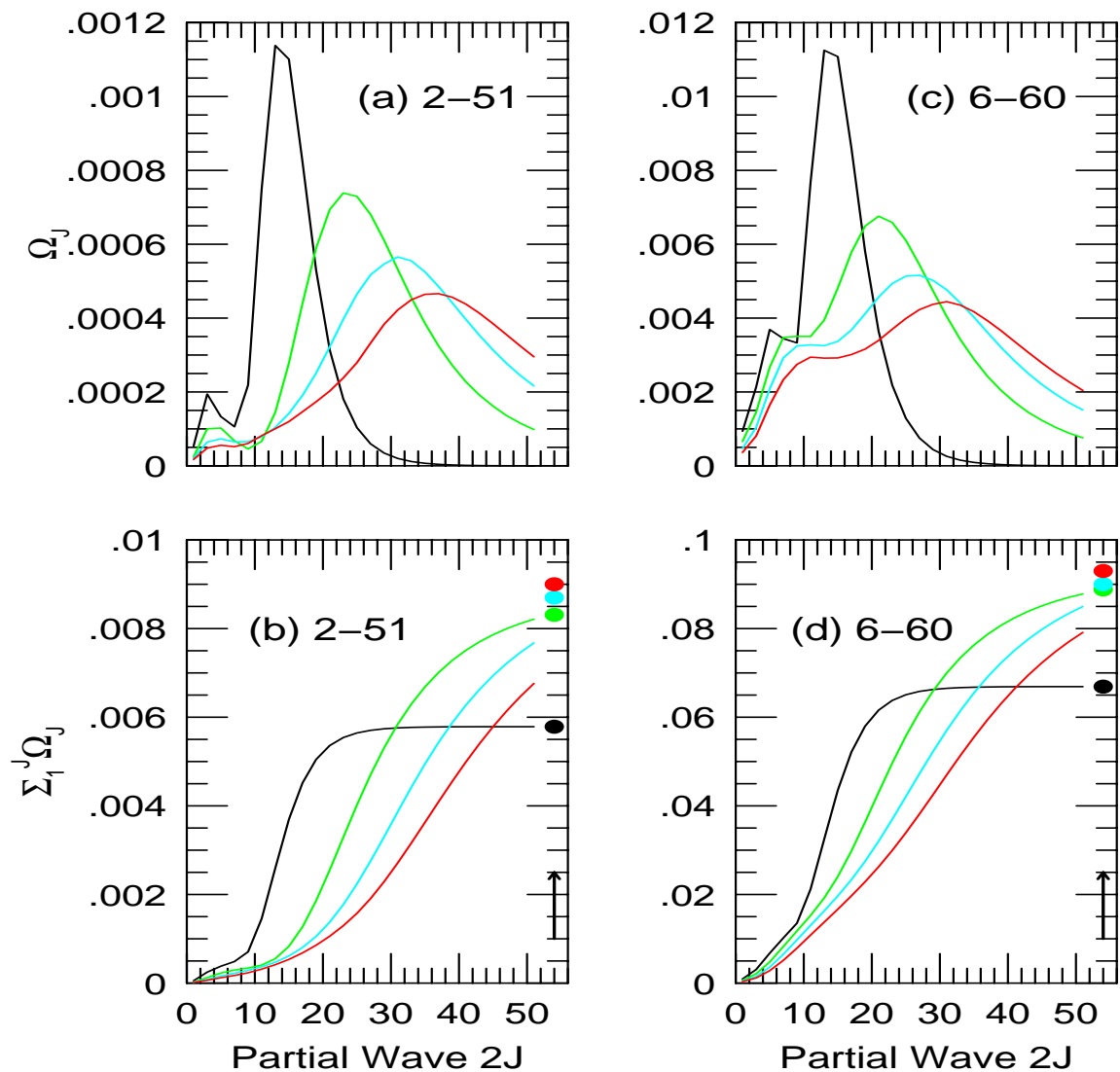,height=17.0cm,width=18.0cm}
\caption{
Partial wave (PW)
collision strength $\Omega_J$ and partial sum of PW collision strength
$\sum_{1/2}^J\Omega_J$ for E2 transitions between excited levels 2-51 and 6-60.
Other symbols are the same as Figs.~3.
%Notations are the same as Figs.~3.
}
\end{figure}

\begin{figure}
\centering
\psfig{figure=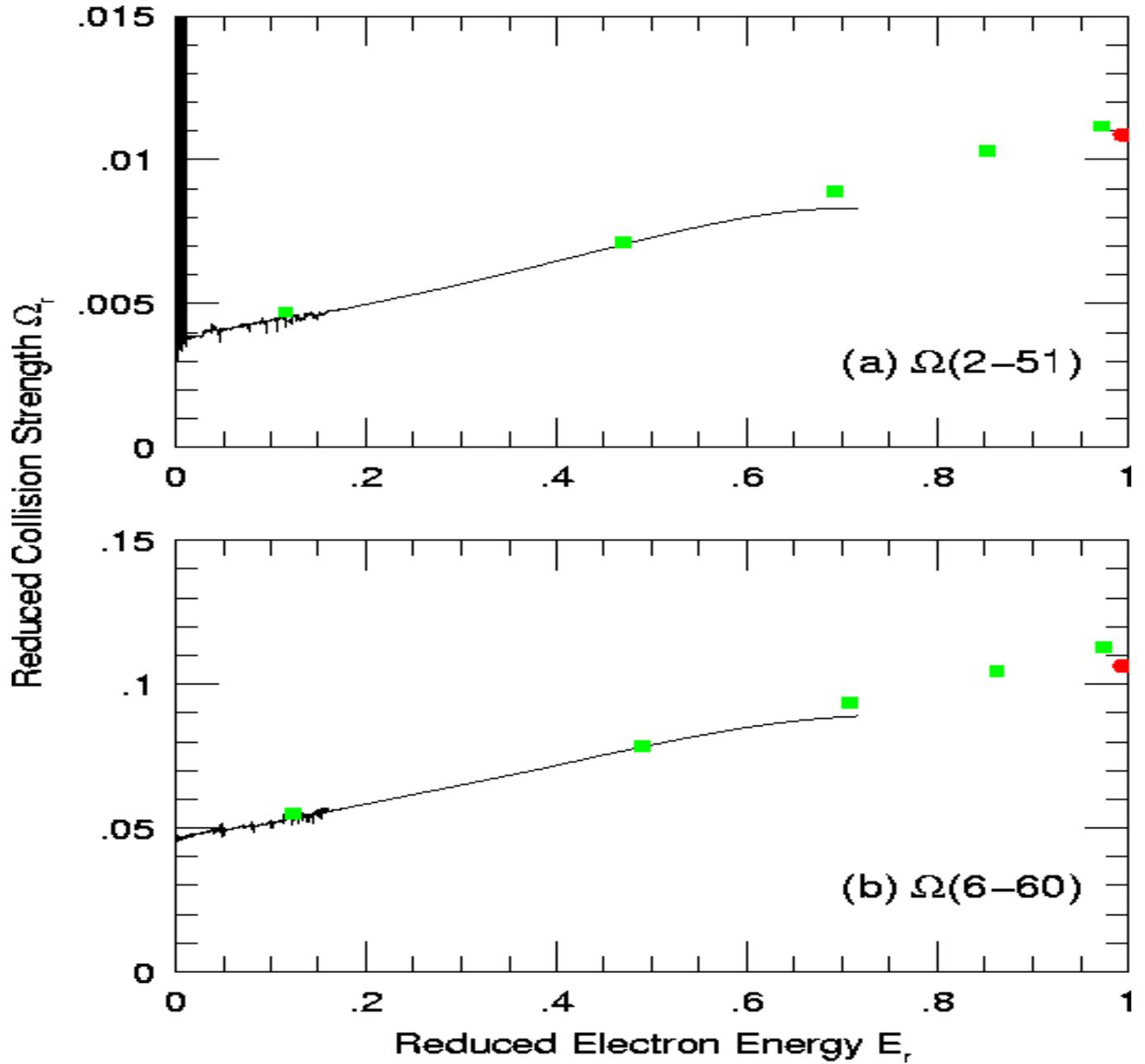,height=17.0cm,width=18.0cm}
\caption{
Reduced collision strengths $\Omega_r$
are plotted as a function of reduced electron
energy $E_r=E_i/(E_i+c)$ (c=e is a constant, $E_r\in$ [0,1]) for E2 transitions 2-51
and 6-60. The infinite energy red filled circles
and our RDW blue squares are also shown for comparisons. 
}
\end{figure}

\begin{figure}
\centering
\psfig{figure=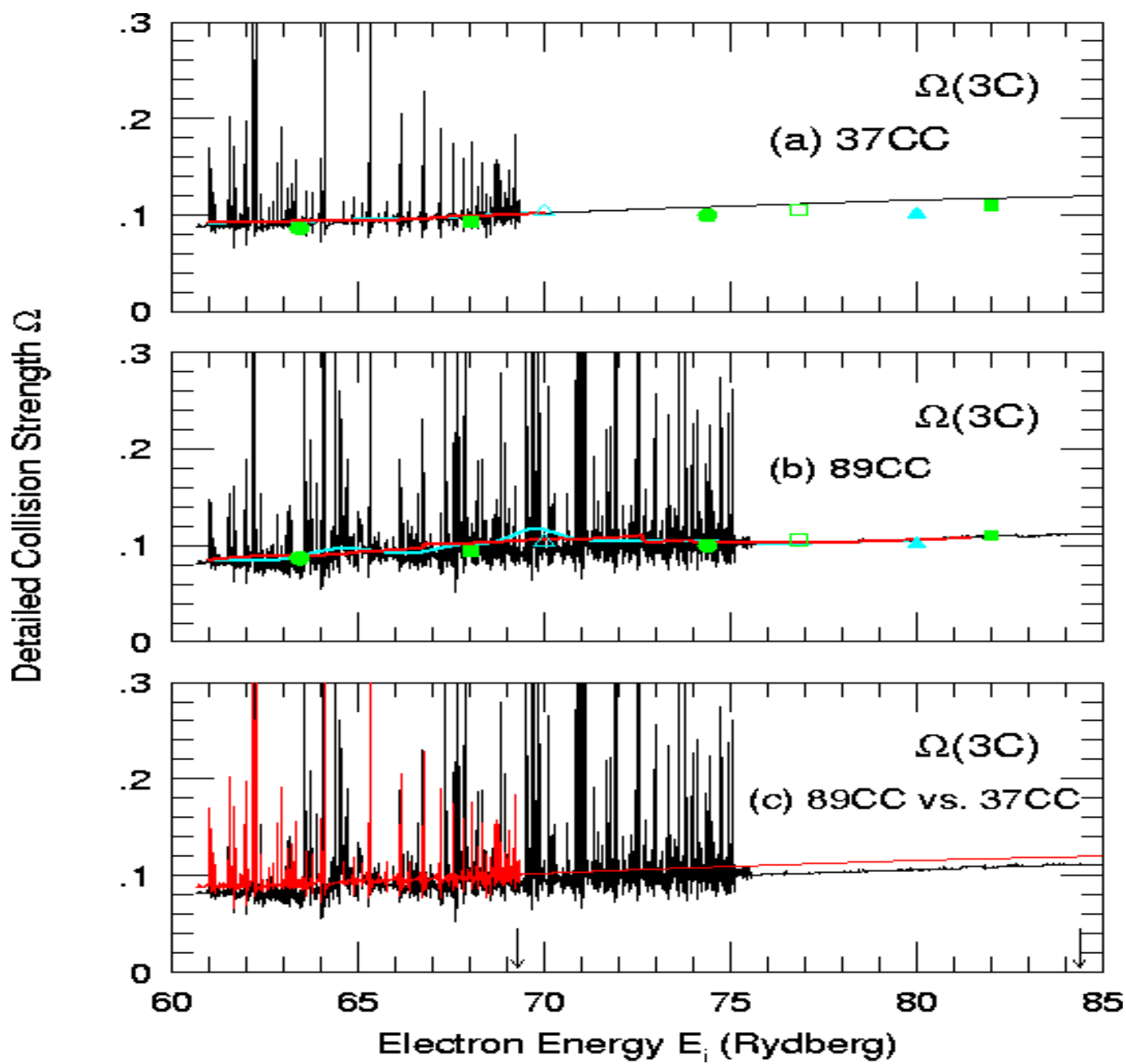,height=17.0cm,width=18.0cm}
\caption{
Comparisons of collision strength $\Omega$ with detailed
resonance structures versus incident electron energy $E_i$ between 37CC and
89CC BPRM calculations for dipole allowed E1 transition 3C (1-27, spin unchange).
(a) 37CC; (b) 89CC; (c) 37CC (red curve) vs. 89CC.
In (a) and (b), the green and red dashed-lines are the numerical averaged (NA) and
gaussian averaged (GA) collision strengths, respectively; the green dots, filled squares,
open squares and filled and open blue triangles are RDW values \cite{zh89}, RDW values \cite{cgx},
DW values \cite{bh92} and previous R-matrix values \cite{mo97} and \cite{gu00}, respectively.
The arrows in (c) represent 37CC and 89CC thresholds.
}
\end{figure}

\begin{figure}
\centering
\psfig{figure=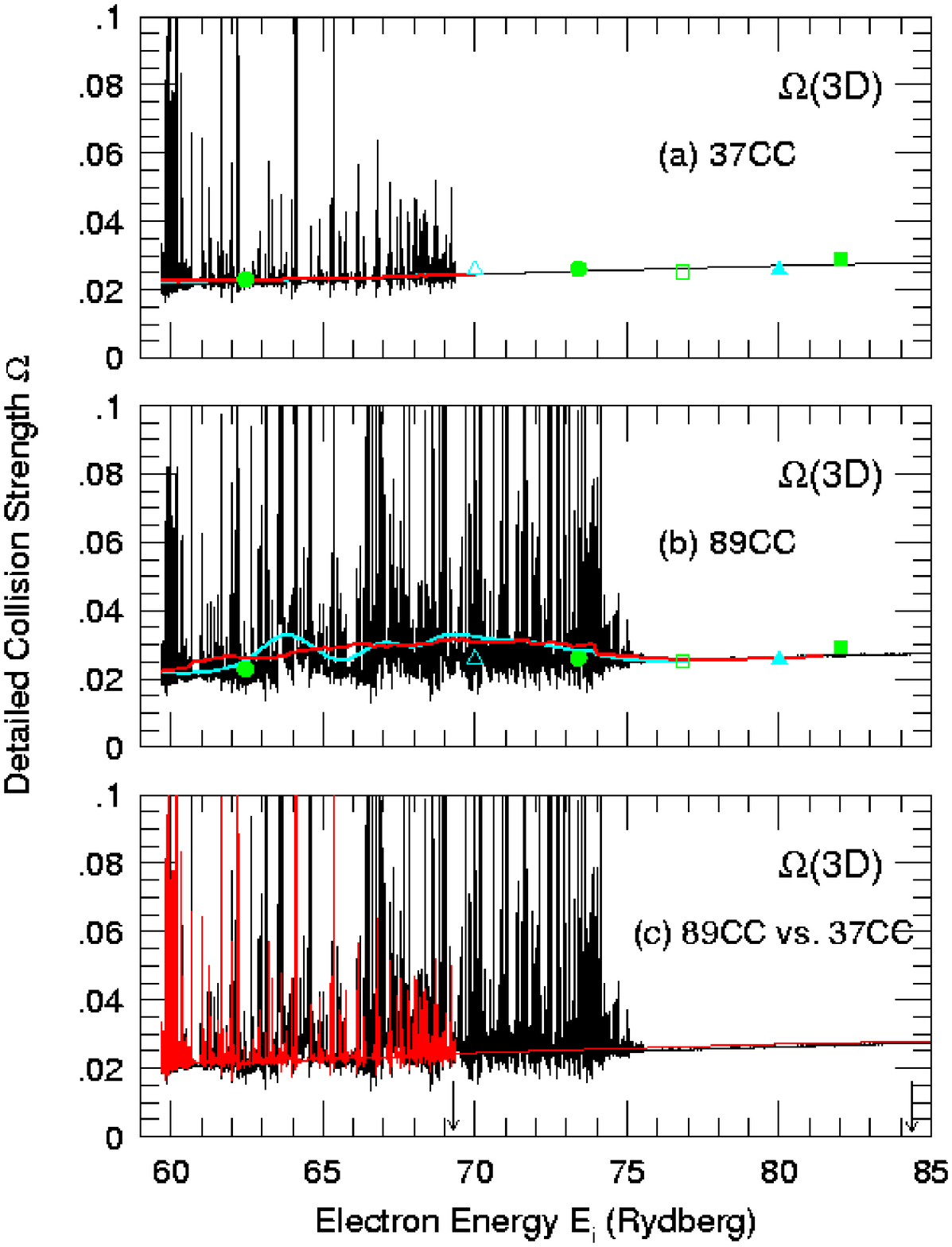,height=17.0cm,width=18.0cm}
\caption{
Comparisons of collision strength $\Omega$ with detailed
resonance structures versus incident electron energy $E_i$ between 37CC and
89CC BPRM calculations for intercombination transition (dipole allowed spin-changed)
E1 transition 3D (1-23). The other symbols are the same as Figs.~7.
}
\end{figure}

\begin{figure}
\centering
\psfig{figure=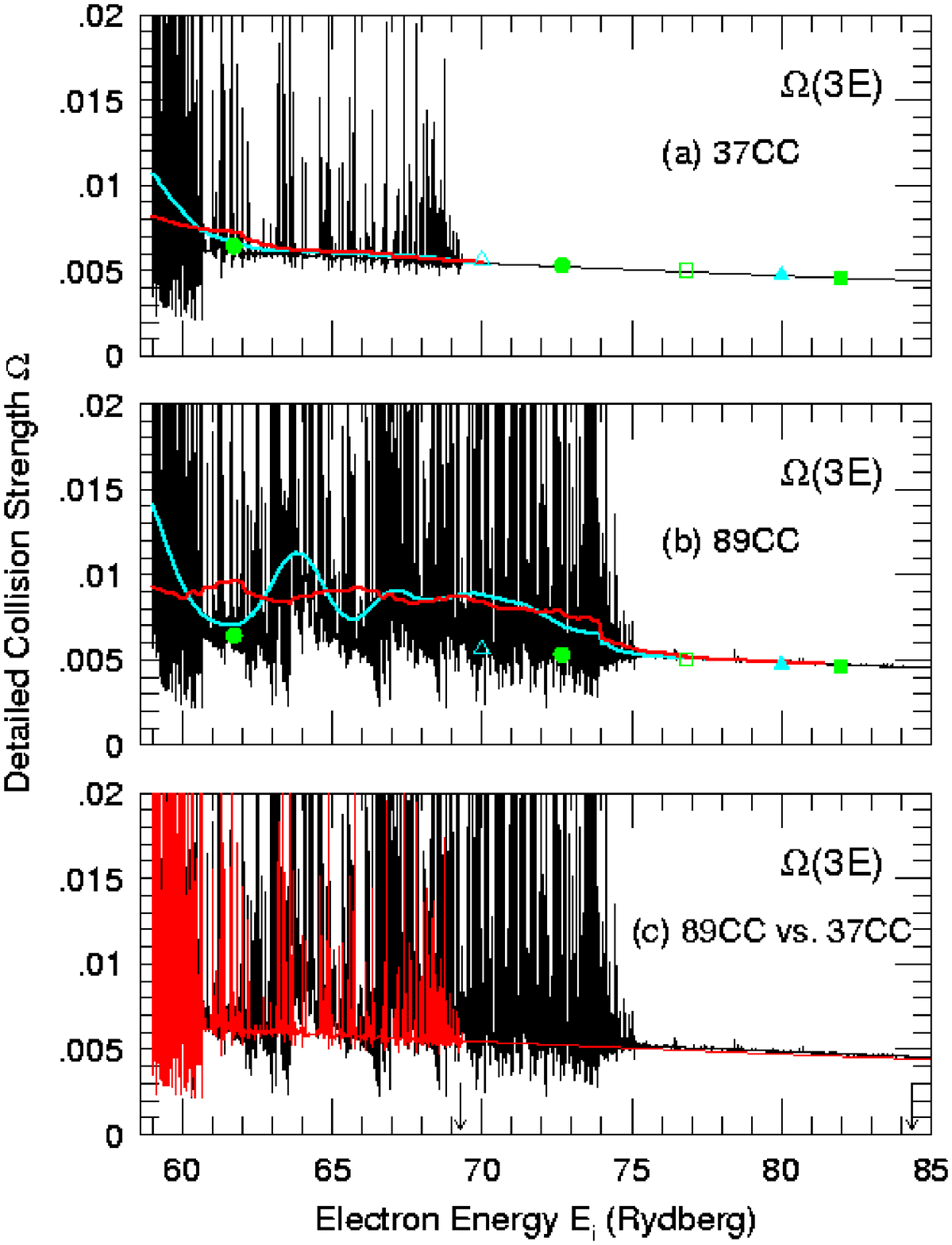,height=17.0cm,width=18.0cm}
\caption{
Comparisons of collision strength $\Omega$ with detailed
resonance structures versus incident electron energy $E_i$ between 37CC and
89CC BPRM calculations for intercombination transition (dipole allowed spin-changed)
E1 transition 3E (1-17). The other symbols are the same as Figs.~7.
}
\end{figure}

\begin{figure}
\centering
\psfig{figure=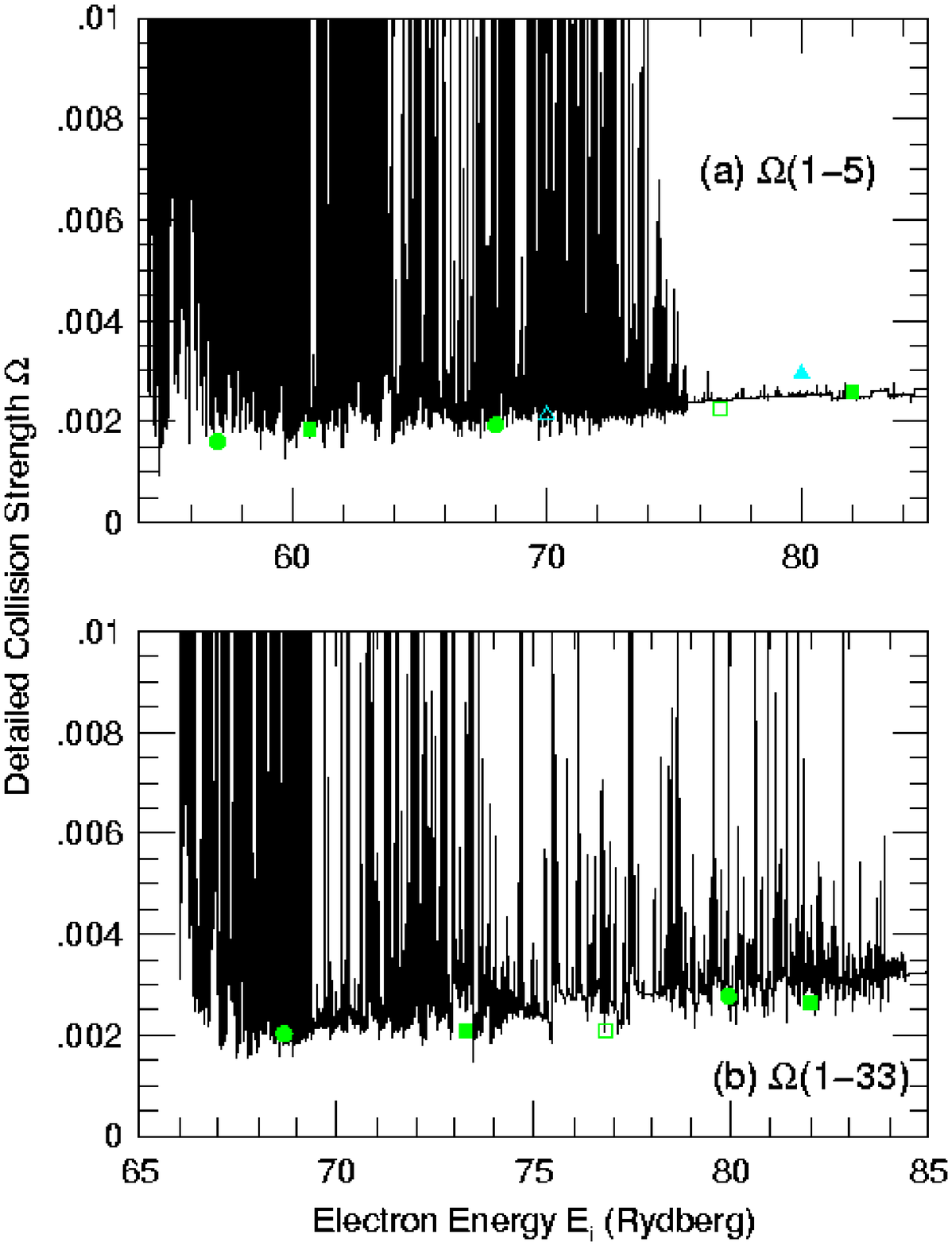,height=17.0cm,width=18.0cm}
\caption{
Collision strength $\Omega$ from 89CC BPRM calculations
with detailed resonance structures as a function of incident electron energy $E_i$
for pure E1 transitions from the ground state: (a) 1-5; (b) 1-33.
The other symbols are the same as Figs.~7. 
}
\end{figure}

\begin{figure}
\centering
\psfig{figure=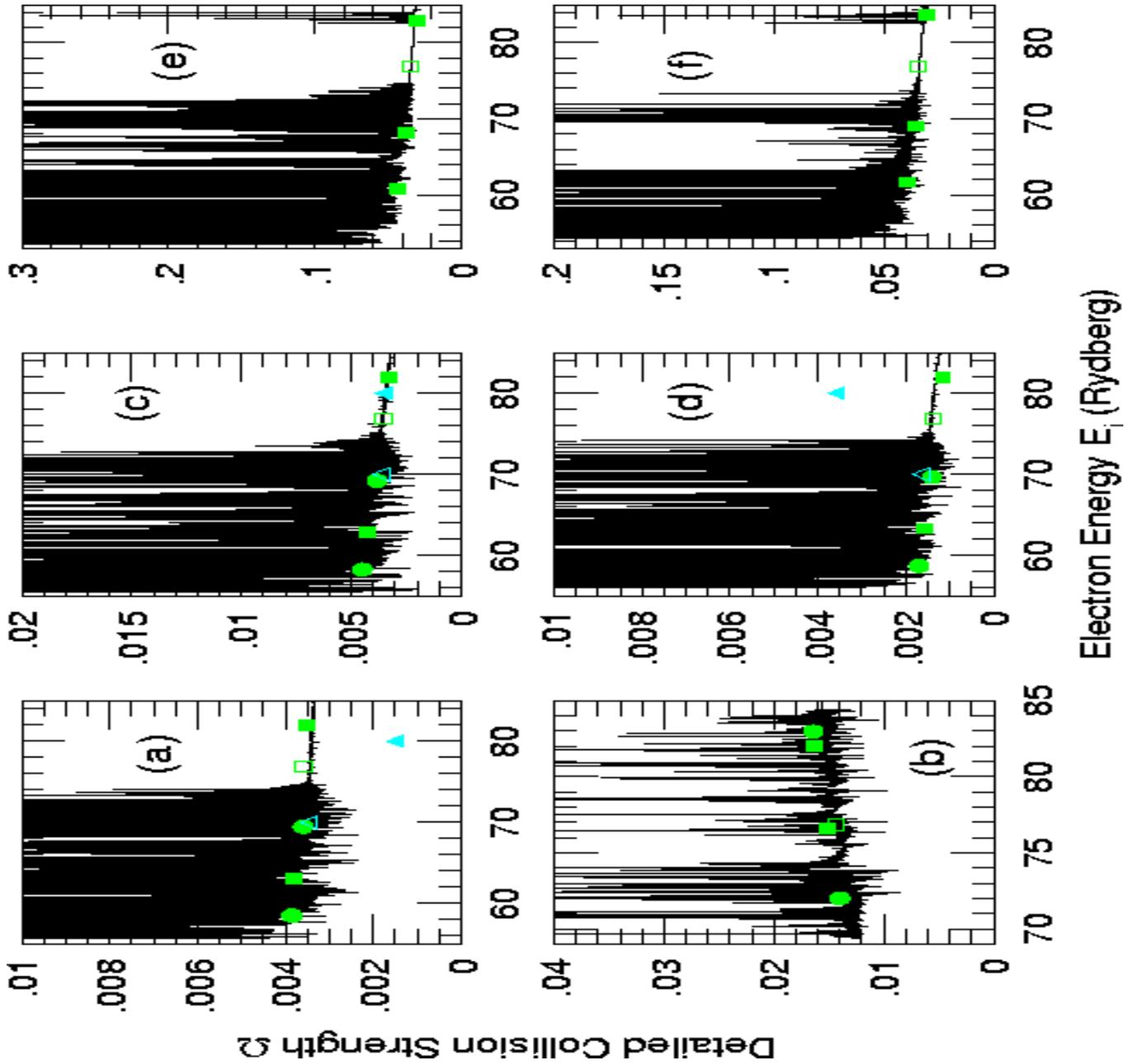,height=17.0cm,width=18.0cm}
\caption{
Collision strength $\Omega$ from 89CC BPRM calculations
with detailed resonance structures as a function of incident electron energy $E_i$
for E2 and M1 transitions: (a) pure E2 (1-7); (b) pure E2 (1-37); (c) pure M1 (1-6);
(d) pure M1 (1-9); (e) mixed E2+M1 (2-3); (f) mixed E2+M1 (2-5).
The other symbols are the same as Figs.~7. 
}
\end{figure}

\begin{figure}
\centering
\psfig{figure=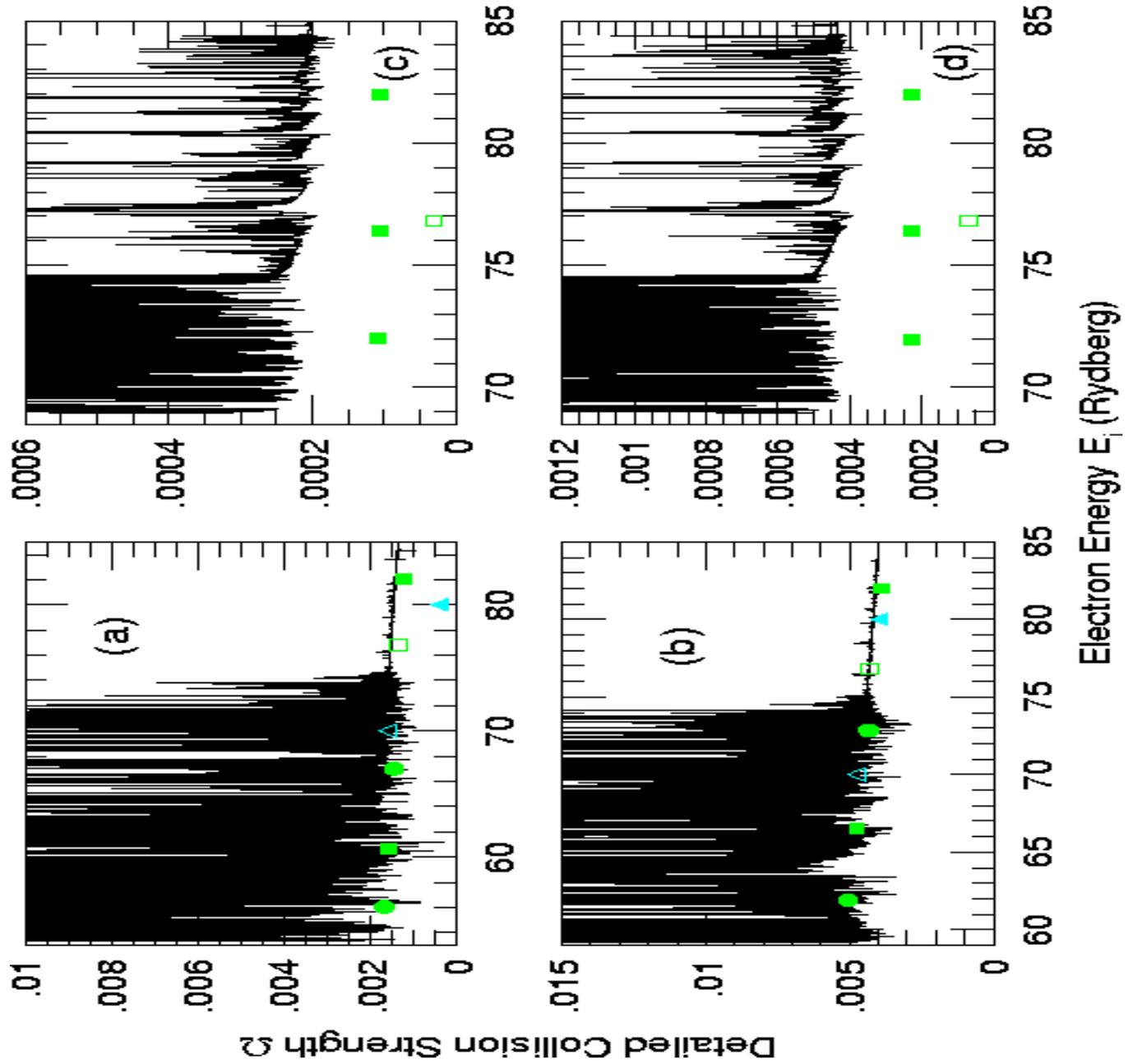,height=17.0cm,width=18.0cm}
\caption{
Collision strength $\Omega$ from 89CC BPRM calculations
with detailed resonance structures as a function of incident electron energy $E_i$
for E1, M2 and E3 transitions: (a) pure M2 (1-2); (b) pure E3 (1-20); (c) mixed
M2+E3 (3-36); (d) mixed E1+M2+E3 (2-35).
The other symbols are the same as Figs.~7. 
}
\end{figure}

\begin{figure}
\centering
\psfig{figure=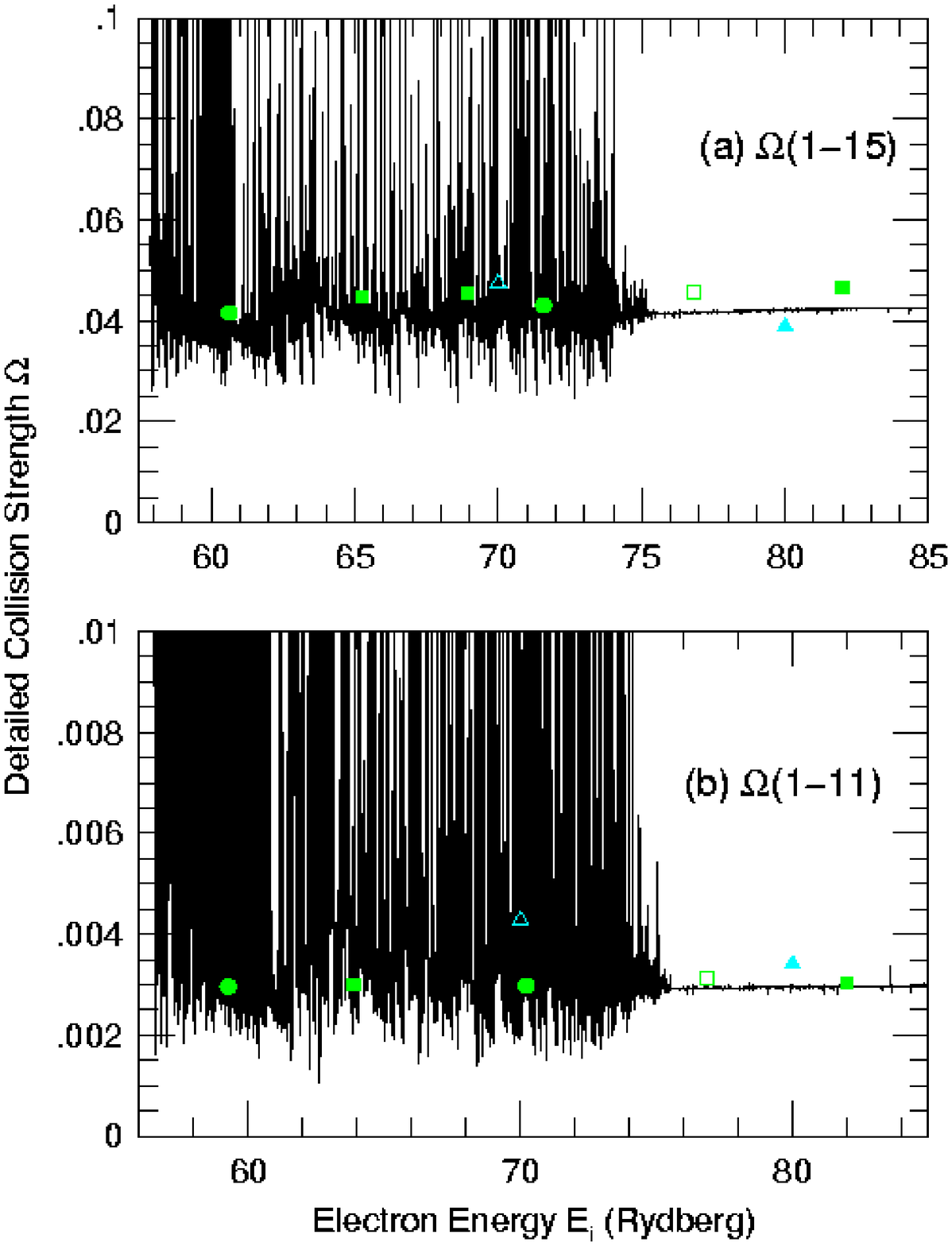,height=17.0cm,width=18.0cm}
\caption{
Collision strength $\Omega$ from 89CC BPRM calculations
with detailed resonance structures as a function of incident electron energy $E_i$
for monopole transitions from the ground state: (a) 1-15; (b) 1-11.
The other symbols are the same as Figs.~7. 
}
\end{figure}

\begin{figure}
\centering
\psfig{figure=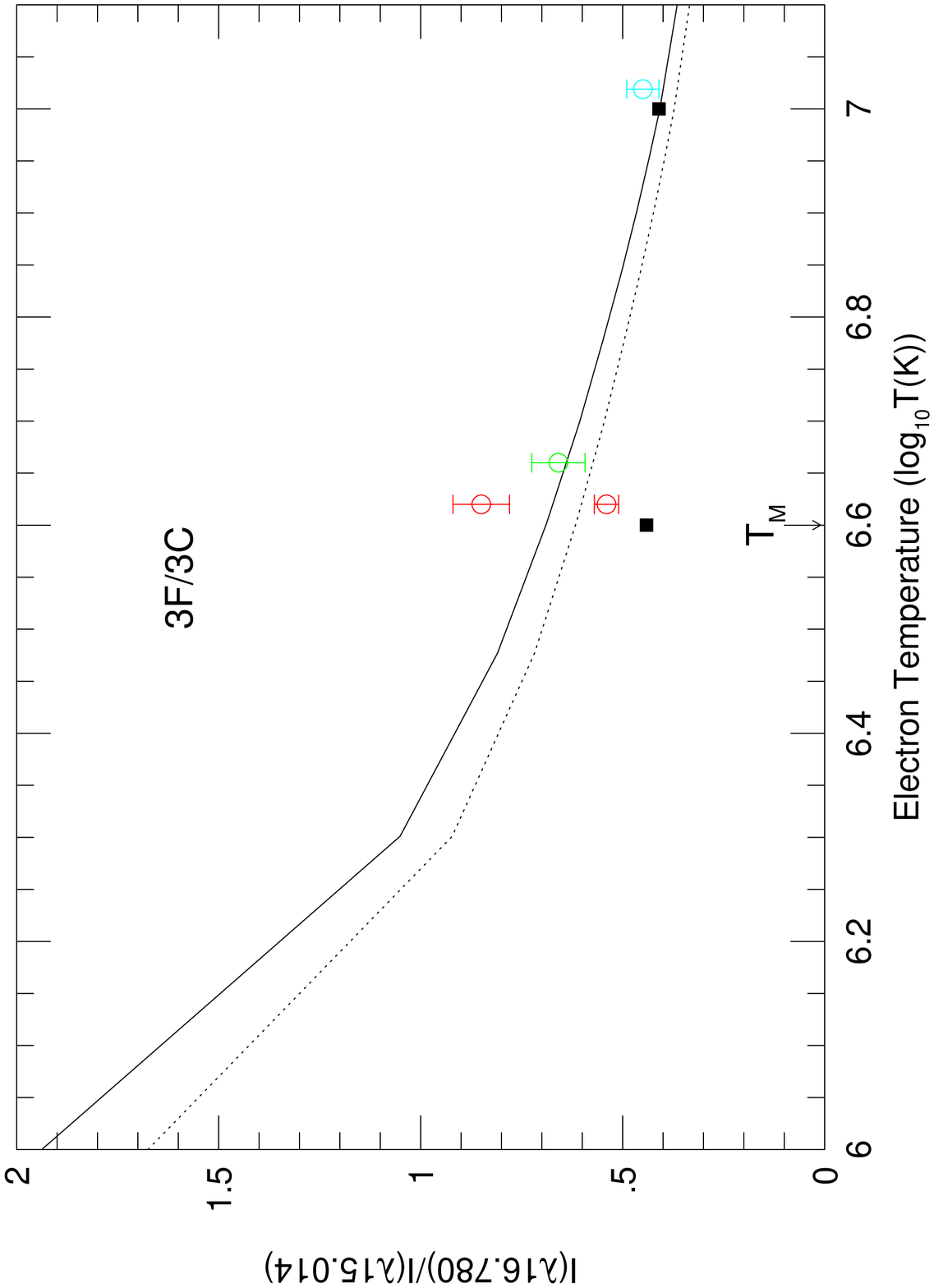,height=17.0cm,width=18.0cm}
\caption{
Intensity ratios of two {\small X}-ray lines 3F/3C (1-5/1-27)
from \Fe17 as a function of electron temperature, calculated at
two electron densities 10$^{13}$ cm$^{-3}$ (solid lines) and 10$^9$ cm$^{-3}$ (dotted lines),
and compared with observed and experimental values:
%The 4 open circles with error bars are observed and experimental values:
from the solar corona
at T$_m \sim$~4 million degrees Kelvin \cite{hu76}, %(Hutcheon \etal 1976),
from the corona of solar-type binary star Capella at $\sim$~5 million degrees Kelvin \cite{ca00},
%(Canizares \etal 2000),
and from the EBIT experiment at $\sim$~10 million degrees Kelvin \cite{la00}. %(Laming \etal 2000).
The filled squares are values using previous cross sections \cite{bh92}
%(Bhatia and Doschek 1992)
which differ from observations at low temperatures.
}
\end{figure}

%***
%***  E n d   o f   p a g e   6   o f   g a l l e y - m o d e   o u t p u t
%***


\begin{thebibliography}{}
\bibitem{not1}The calculations and collision strengths presented in \cite{mo97}
are in fact a background calculations similar to the various distorted-wave
calculations. There are also some other limitations for this calcualtion, e.g.
the convergence of partial-wave expansion (only 2J up to 9 is employed in their
27CC R-matrix calculation). In many literature publications and spectra diagnostics
applications, as this calculation has been extensively referred to 
a representative R-calculations for \eFe17, many confusions have thus been caused.
\bibitem{not2}Although Eq.~(5) in \cite{br98} can be used for DW collision strengths,
this equation is generally wrong due to the presence of complex resonance structures
shown in the present Figures 7-9. In contrast, Eqs.~(4) and (5) in this paper is generally
correct to be applied in EBIT experiments. The CRM used in \cite{la00} is generally
correct, but the collision strengths they used did not include huge resonance
ehancements. Hence, the theoretical intensity ratios reported in \cite{la00} suffered
similar problem to those in \cite{br98}.
\bibitem{not3}75.0 eV is the highest thresthold with strong `resonance' for lines
3C, 3D and 3E. In the same target model, the highest strong `resonance' thresthold
is different for different lines. For example, in Fig.~10(b)
a strong line 1-37 ($1s^22s2p^63d~^1D_2$)
has very strong and dense resonance enhancements over entire region up to 84.5 Ry, the highest
energy threshold of 89-level target model.
\end{thebibliography}
\end{document}